# The Many Computations Interpretation (MCI) of Quantum Mechanics

Dr. Jacques Mallah     September 4, 2007

**Abstract:**

Computationalism provides a framework for understanding how a mathematically describable physical world could give rise to conscious observations without the need for dualism. A criterion is proposed for the implementation of computations by physical systems, which has been a problem for computationalism. Together with an independence criterion for implementations this would allow, in principle, prediction of probabilities for various observations based on counting implementations. Applied to quantum mechanics, this results in a Many Computations Interpretation (MCI), which is an explicit form of the Everett style Many Worlds Interpretation (MWI). Derivation of the Born Rule emerges as the central problem for most realist interpretations of quantum mechanics. If the Born Rule is derived based on computationalism and the wavefunction it would provide strong support for the MWI; but if the Born Rule is shown not to follow from these to an experimentally falsified extent, it would indicate the necessity for either new physics or (more radically) new philosophy of mind.

**Introduction:**

The no-collapse wavefunction-only Many Worlds Interpretation (MWI) of Quantum Mechanics has much to recommend it, especially the simplicity (compared to any other possible interpretation) of not adding additional physics to the quantum wave equation [Everett, Tegmark]. However, proper explanation of the role of the observer is needed, and if the MWI is to be convincing there should be an explanation of how the Born rule for observation probabilities follows from the dynamics of the wavefunction.

The derivation of the Born rule should make clear how the role of the observer relates to the observation probabilities. Also, it would be helpful to know at least in principle how such a derivation could be generalized to generate observation probabilities for observers within other mathematically possible models of a physical world.

It would be better to show that the MWI does positively predict the Born rule for observations, rather than merely ruling out any alternative formulas, because one must take seriously the possibility that the MWI does not in fact predict conscious observations and is therefore not a viable theory. (For example, the advocates of the single-world Pilot Wave Interpretation would have to argue that the wave alone generates no observations, in order to avoid a MWI. [Lewis])

It seems intuitively obvious that probabilities in a many worlds theory are proportional to the number of observers that experience a given outcome. Attempts to derive the Born rule which do not respect this intuition do not seem likely to achieve the desired properties listed above, and have been unconvincing [Van Esch].

To address such issues, a precise framework in which observers can be included in a picture of a physical world is needed. Such a framework may be offered by computationalist philosophy of mind, in which the computations performed by a physical system account for the conscious observations made in that system. The application of computationalism to the MWI was proposed by Chalmers [Chalmers 1] among others, but important details have been lacking.

It is important to note that, besides for having potential to clarify the MWI, computationalism must be taken seriously in its own right and <u>those who would reject it then must still face the question of under what circumstances matter would give rise to conscious thought</u>. Those who refuse to take a position on the matter and simply assume that the MWI (or any other model of physics) would work should be under no illusion that they obtain a simpler theory by doing so; in fact they propose no theory whatsoever regarding observations.

**Advantages of the computationalist approach:**

1) Computationalism is independent of quantum mechanics in terms of its motivation and history. It would apply equally to any model of mechanics within a wide range, at least for those that can be cast as initial-value problems. This range includes classical particle mechanics, wave mechanics, quantum mechanics, QM with nonlinear modifications, QM with hidden variables, and digital systems. The concept of computation is a way of characterizing the structure and function of any dynamical system.

2)      The computational way of characterizing systems should be considered to be part of *philosophy of math*.  With that understanding, one can distinguish the application of it to the study of physical systems from other schemes that fall strictly into the category of proposals for new psycho-physical laws of nature, such as Page's Mindless Sensationalism [Page 2].  Computationalism would thus allow the study of the role of consciousness in physics without the need for philosophical dualism, although it would be *compatible* with either materialism or dualism [Chalmers 1].

3)      The computationalist approach first reduces the problem of linking the mathematical description of the physical world to consciousness to the (hopefully) easier and more comprehensible problem of linking the mathematical description of the physical world to computation.  The second half of the problem, that of finding the particular types of computations that would be conscious and describing their nature, is left to later study by neuroscientists and workers in artificial intelligence, aka the neural correlates of consciousness or the "easy problem" of [Chalmers 1].

These advantages come at a price:

Approaches to linking consciousness to the physical world that introduce proposed new natural laws are free to choose laws that give any desired results.  Thus, it is trivial for them to put in Born's Rule by hand.  If computationalism is to have greater philosophical weight, it must not have that kind of "all parameters are free parameters" licentiousness.

Of course, "general philosophical considerations" are subject to uncertainty and controversy, and there are various options one could take, perhaps with a view towards what the probabilities must end up being.  Indeed, in the face of uncertain intuitions and disagreements, looking to nature as a referee makes sense.  The plausibility advantage of computationalism therefore is merely one of degree, and each option within computationalism requires supporting arguments (independent of experimentally obtained knowledge) to make it seem plausible.

**Why is derivation of the Born Rule important?**

If the Born Rule can not be derived from the wavefunction dynamics, it could be added as an additional postulate, as has sometimes been suggested.  Yet advocates of the MWI have been loathe to take such a step.  There are three reasons why actually *deriving* the Born Rule is considered important:

1)      It is <u>necessary</u> to prove that the standard wavefunction dynamics need <u>not</u> imply some *other* rule that has been experimentally falsified.  In particular, the rule in which each outcome recieves equal probability is often thought to be more natural from the point of view of wavefunction dynamics.  If the wavefunction dynamics do demand the wrong rule, it would block adding the Born Rule as an additional postulate since the wrong probability rule is already derived from the dynamics; the MWI would be falsified.

2) The Born Rule is all about probabilities of conscious observations. Because the actual physical dynamics would not be changed, it is far from clear that it is even possible to add the Born Rule as an additional postulate without it becoming a manifestly dualist model of physics in which the Born Rule is a psycho-physical law relating non-physical consciousness to the physical world. Besides for being controversial and against the intuitions of most MWI advocates, such a dualist approach would inevitably open the door to other psycho-physical laws, and thus to the wholesale addition of new laws to the the theory of nature, a far cry from the attractively spartan nature of the original MWI.

3) Adding an additional postulate, even if it could be construed as not dualist, would make the MWI more complicated. This would decrease the advantage the MWI holds over rival interpretations in terms of its Occam's Razor simplicity.

**Why existing 'derivations' of the Born Rule for the MWI fail:**

There is a fundamental reason why no mathematical derivation could *really* derive any predictions of observation probabilities without a theory of consciousness: Observation probabilities are by definition about conscious observations. Only computationalism seems like it might be able to bridge the gap between consciousness and mathematics without adding psycho-physical laws to the model.

The question then arises: Why does the problem not crop up with classical mechanics? The truth is, it *does*. Without some sort of theory of consciousness, there are no grounds to say that a human brain is any more conscious than a rock, and certainly no rule to give the probabilities of whatever conscious observations might be generated in these rocks. Even for classical mechanics, computationalism faces the Implementation Problem to be discussed in the next section, for which a solution is proposed in this paper.

Still, in classical mechanics it is much easier to sweep the problem under the rug, because there is basically only one relevant outcome for a given experiment. It is still possible to ask questions about probabilities of conscious experiences, such as: What is the probability of being female? Female brains work a bit differently than male brains, so they might on average generate a different number of conscious experiences per unit time, giving a bias to the probability that a simple population count would not reveal. But the only experiment one can do is to look in the mirror to check one's own gender. Once that's done, the experiment cannot be repeated to gather more data; the one answer that applies to you is already known. Even if the real probability is 75%, a single experiment per person leaves one with little data to go on.

With Many Worlds QM, every quantum experiment puts one into a similar situation to the male/female check, but with the key difference that there are many such similar experiments that one can do. These experiments have (for most observers) shown a pattern, summarized by Born's Rule, that needs explaining.

If probabilities are proportional to the number of "observers" that inhabit each branch, which is certainly concievable even though it would require much more specification to evaluate (such as identifying "observers" with implementations of computations), then clearly any attempted derivation of Born's Rule which does not count the observers properly is not correct, unless it proves that the *only* way in which observers *could* be distributed must match the Born Rule.

Even then such a 'derivation' would only have done half the job; it must still be shown (without appeal to experiment) that observers *would* exist in such a mathematical model. (The advocates of the Pilot Wave Interpretation would be delighted if only particles and not waves could give rise to observers, for example.)

There have been many attempts to derive the Born Rule for the Many Worlds Interpretation, which is quantum mechanics without collapse, hidden variables, or other modifications such as new laws of nature. To show that these derivation attempts fail even at the first half of the job, it is sufficient to produce a counter example: an alternative (though perhaps *experimentally* disproved) possibility for observation probabilities which the attempted derivations don't disprove.

The Alternative Projection Postulate (APP) of [Van Esch] provides just such a counterexample. Boiled down to its essentials, the APP implies that each possible type of observation in different branches has the same probability. These may be associated with brain states with nonzero amplitude. Possible observations don't include the ones that would be associated with branches of zero amplitude, but otherwise the actual amplitudes don't matter. For example, if the wave function of the universe is

$|\Psi\rangle = [ |A\rangle (|1\rangle + |2\rangle) + |B\rangle|3\rangle ] / 3^{½}$

and "A" and "B" are possible observations, then A and B have the same probability.

A generalized APP (GAPP) could include observers that overlap in terms of their branches. For example, if the wavefunction is

$|\Psi\rangle = |\text{Mork sees red}\rangle (|\text{Mindy sees green}\rangle + |\text{Mindy sees blue}\rangle) / 2^{½}$

then with the GAPP the probability for an observer to be "Mork seeing red" is 1/3, the probability for an observer to be "Mindy seeing green" is 1/3, and the probability for an observer to be "Mindy seeing blue" is 1/3.

The GAPP can be further modified. Instead of having equal probabilities, each possible conscious observation C could have its own intrinsic measure M(C), with probabilities being equal to the measure of that observation divided by the total measure. (The GAPP is the special case M(C) = M(C') for any C,C'. Total measure need not be conserved, as shown by the example of human birth or death.) Therefore, even symmetrical looking wavefunctions could give rise to unequal probabilities. This modification also allows the probabilities to remain normalized even if there are infinitely many possible conscious experiences.

Another modification of the Born Rule leads to a formula that is not only viable, but for a special case is actually believed in by almost everyone who claims to believe the Born Rule: The measure of a 'world' state on a 'branch' of the wavefunction is proportional to the squared amplitude of the branch multiplied by a factor L that depends on the details of the observer states. Call this the Generalized Born Rule (GBR).

For example, if $|\Psi\rangle = a\ |A\rangle|B\rangle + b\ |C\rangle|D\rangle$, note that $L(|A\rangle)$ need not equal $L(|B\rangle)$; the measure $M(\text{"A"}) = L(|A\rangle)\ |a|^2$.

In the special case with L = "the number of macroscopically distinguishable sentient *lives*" in the state, this is easily believed. For example, if no one can live in a branch, it will not be observed; that is known as the Anthropic Principle and helps explain the fine-tuned nature of the physical 'constants' for our universe. [Tegmark]

Any purported proof of the Born Rule had better *not* rule out this case of the GBR, which I will call the Anthropic Born Rule (ABR). It is the ABR that physicists really believe and that has experimental backing. *When speaking about a proof of the Born Rule, what should really be understood is that the ABR is the rule that needs proving.* This is an important distinction because measure, unlike "probability", is not conserved.

The trick that makes the general GBR important is that the factor L could be a more complicated function of the physical state; for example, there could be a number of *microscopically countable* observers that *differs from* the obvious macroscopic count. For example, the number of neural connections in a brain could concievably be relevant, and that could be influenced by the thought processes of a person who learns the result of an experiment, e.g. judging it a "boring' or 'interesting' result.

[The computationalist approach based on counting implementations could be thought of as similar to the idea of counting "microscopic lives", except that its counting process (being of general nature applicable to any mechanics) assumes a uniform measure per implementation rather than an intrinsic amplitude-squared starting point, if the number of implementations is finite. If the number of implementations is infinite, it could become legitimate to invoke symmetries of the system to determine implementation density ratios, so some of the traditional approaches might make contact with the eventual form of the computationalist program.]

A brief (and necessarily incomplete) critical look at the approaches attempted for derivations is now in order. While not proofs of the Born Rule for the MWI, each of these approaches provides a valuable contribution towards the understanding of quantum mechanics.

- Non-contextuality

Noncontextuality of probabilities means that the probability associated with a component of the normalized wavefunction does not depend on what other orthogonal terms are present. Gleason's Theorem, proven in 1957, shows that only Born's Rule (that the probabilities are proportional to the squared amplitude) is consistent with general noncontextuality of probabilities.

While an impressive result, the catch is that noncontextuality of probabilities has not been derived from the physical model of the MWI. It can be assumed, but going this route is only an advance if it is more plausible *a priori* to assume noncontextuality of probabilities than it would be to directly assume Born's Rule.

Unfortunately, noncontextuality of probabilities is not particularly plausible as an assumption. Consider the Generalized Born's Rule (GBR) mentioned above. L-factors could change as a function of time differently in different branches, providing "contextual effects on probability"; for example if

$(|A> + |B> + |C>)/3^{1/2}$ evolves to $(|A> + |B> + |0>)/3^{1/2}$

where $L(|A>) = 1$, $L(|B>) = 1$, $L(|C>) = 1$, and $L(|0>) = 0$. This increases the probability of $|A>$ from 1/3 before to ½ after the time interval.

It does not, however, influence the measures of $|A>$ or $|B>$, so the conditional probability of $|A>$ given that the state is not $|C>$ remains the same (½). Therefore it cannot be used for signalling across branches, which is fortunate because this example is simply the mundane case in which observer "C" dies. It illustrates that measure is the quantity with real significance, while "probability" is a mere derived quantity.

Conservation of measure is needed for noncontextuality of probabilities and thus for Gleason's Theorem. Measure need not, in the general case, be conserved. *As a result, traditional Gleason-based approaches to the Born Rule should not be relied upon*.

What about noncontextuality of *measure*? This would mean that the measure associated with a component of the normalized wavefunction does not depend on what other orthogonal terms are present. If death is possible it is easy to see that the general case even of this less restrictive assumption *must be false*:

$|\Psi_1> = (|Dead> + |Alive>) / 2^{1/2}$

$|\Psi_2> = (|Dead> - |Alive>) / 2^{1/2}$

$|\Psi_3> = (|\Psi_1> + |Dead2>) / 2^{1/2} = ½ |Alive> + ½ |Dead> + (|Dead2> / 2^{1/2})$

$|\Psi_4> = (|\Psi_1> + |\Psi_2>) / 2^{1/2} = |Dead>$

|Dead>, |Alive>, and |Dead2> are mutually orthogonal, so $<\Psi_1| \Psi_2> = 0$. Assume that $M(½ |Alive>) > 0$, so $M(|\Psi_3>) > 0$. $M(a |Dead>) = M (b |Dead2>) = 0$ for any a,b.

Measure must be greater than or equal to zero, so if noncontextuality of measure were true, changing $|\Psi_3>$ into $|\Psi_4>$ by substituting $|\Psi_2>$ in place of |Dead2> could not decrease the total measure (since both of those are orthogonal to $|\Psi_1>$). Yet clearly $M(|\Psi_4>) = 0$, while $M(|\Psi_3>) > 0$. Note that if we were dealing with a conserved probability, rather than measure which need not be conserved, this proof would not apply since the total "probability" of any normalized wavefunction is traditionally always assumed to be 1.

Measure can't be non-contextual in the general case, but measure could still be "partly non-contextual" if additional restrictions are imposed; for example, that each orthogonal term must be written in a preferred basis. The GBR satisfies this assumption, but so do other formulas. It may sound plausible that there is no preferred basis, but on closer examination that is far from clear; for example, energy eigenstates always constitute a sort of special basis. More importantly, the concept of "brain states" implies some sort of preferred basis already. Implementation of computations seems to require a simple sort of preferred basis as well, which I will take to be the position basis.

Only an argument that measure is proportional to a wavefunction-dependent but non-contextual *something* with a conserved sum, multiplied only by the obvious macroscopic L-factors, could legitimize a Gleason-based approach to the ABR. That *something* would need to be identified however, and arguments made in support of its conservation and its non-contextuality.

Consider the wavefunctions

$|\Psi_1> = $ a |A> |1> + b |A>|2> + d |C>|3> = c |A>|D> + d |C>|3>

$|\Psi_2> = $ a |A>|1> + b |B>|2> + d |C>|3>

|1>, |2>, and |3> are mutually orthogonal. Assume they give rise to no observers, but |A>, |B>, and |C> each give rise to one observer.

According to the Alternate Projection Postulate (APP), the probability of "C" would be ½ in the case of $|\Psi_1>$, but 1/3 in the case of $|\Psi_2>$. This explicitly violates non-contextuality of probabilities, which of course is acceptable. The APP also implies that the measure of "A" is the same in both cases, violating non-contextuality of measure.

Partial non-contextuality of measure could be incorporated into a Modified APP (MAPP) by letting orthogonal relative states in some preferred basis differentiate observers, so that for example in $|\Psi_1>$, the probability of "A" would be 2/3. Note that without a preferred basis, the number of orthogonal components of such relative states would not be defined; |D> (one state) would serve as well as |1> and |2> (two states).

- Locality

Locality of measure means that events in space-like seperated regions from an observer cannot influence the measure of that observer. If the assumption is also made that measure is always conserved as a function of time, this implies that the probability for the observer cannot be influenced. Because measure is not conserved in general, a more general analysis is useful.

Assume that the wavefunction is of the form $|\Psi\rangle = a |A\rangle|E_A\rangle + \ldots$ where there are no other terms that include $|A\rangle$ or interfere with it. If $|A\rangle$ and $|E_A\rangle$ are space-like seperated, the restrictions imposed by the assumption of locality of measure can then be written as $M_A = f(a, |A\rangle)$, where $M_A$ is the measure due to the $|A\rangle$ part of the wavefunction. (It is possible that $|E_A\rangle$ gives rise to measure of its own, which may be different.)

This implies that time evolution of $|E_A\rangle$ into $|E'_A\rangle$ will not affect the measure $M_A$. Born's Rule obviously satifies this, since $\langle E_A|E_A\rangle = \langle E'_A|E'_A\rangle$; for example if $|E_A\rangle = (c(t) |1\rangle + d(t) |2\rangle)$, then $|c(t)|^2 + |d(t)|^2 = 1$, and so $|a(t)|^2 = |c(t)|^2 + |d(t)|^2$ is conserved.

It is clear that some possibilities *are* ruled out by locality of measure; for example, the MAPP defined above is inconsistent with locality of measure since the measure would change under the MAPP if c(t) becomes zero. A similar problem will arise with some possibilities for computationalism (MCII#1, defined later, in the Schrodinger picture).

However, any function of the amplitude is consistent with locality; in particular, if brain state $|A\rangle$ is taken to include the position of a brain (so that $|E_A\rangle$ is spacelike seperated from $|A\rangle$, and different terms' "$|A\rangle$" are not so seperated), letting $M_A = f(a, |A\rangle) = \{1$ if $|a| > 0$, else $0$) yields a kind of APP.

If locality of measure were violated it could allow FTL signalling, which would require a preferred reference frame to determine simultaneous events. *Yet the Shrodinger picture does come equipped with a global time and thus a preferred reference frame.* Might it be proven directly from the dynamics that nonlocal effects *on measure* could not occur if that picture were true? If not, one might resort to the Heisenberg picture which can be made explicitly local [Rubin], but then these pictures would not be equivalent. Most attempts to derive the Born Rule have been presented in the Shrodinger picture. It is true that the Shrodinger picture is ill-defined in quantum field theory due to infinite renormalization, but most physicists believe that infinite renormalization is unphysical and will not be a feature of quantum gravity, which provides the Plank length as a finite scale relevant to renormalization.

- Eigenstates of the frequency operator for infinite ensembles

The frequency operator is the operator associated with observable that is the number of cases in a series of experiments that a particular result occurs, divided by the total number of experiments. [Caves] If is assumed that just the frequency itself is measured, and if the limit of the number of experiments is taken to infinity, the eigenvalue of this frequency operator is unique and equal to the Born Rule probability. The quantum system is then left in the eigenstate with that frequency; all other terms have zero amplitude, as shown by Finkelstein and Hartle.

This scheme is irrelevant for two reasons. First, an infinite number of experiments can never be performed. As a result, terms of all possible frequencies remain in the superposition. Unless the Born Rule is assumed, there is no reason to discard branches of small amplitude. Assuming that they just disappear is equivalent to assuming collapse of the wavefunction.

Second, in real experiments, individual outcomes are recorded as well as the overall frequency. As a result, there are many branches with the same frequency and the amplitude of any one branch tends towards zero as the number of experiments is increased. If one discards branches that approach zero amplitude in the limit of infinite experiments, then all branches should be discarded. Furthermore, prior to taking the infinite limit, the very largest individual branch is the one where the highest amplitude outcome of each individual experiment occurred, if there is one.

- Decoherence and variable course graining

Decoherence is an approximate process, so there is no precise definition of where one "branch of the wavefunction" ends and another begins. It is arbitrary; one could divide a "branch" into finer "branches". Therefore, it is said, probabilities of the branches must be insensitive to how the branches are divided up; the integral of the squared amplitude fits the bill, while directly counting "branches" would not; "branches" have no definite number. [Saunders]

The APP provides a ready example of an alternative which has no sensitivity to how branches are divided up. Another possibility is that there is some non-arbitrary way of assigning the number of relevant 'branches' that is not based solely on decoherence; for example, one could identify 'branches' with local maxima in the absolute value of the wavefunction on configuration space. More logically there could be some way of assigning probabilities to *conscious observations* rather than to branches; counting implementations of computations is an example. Implementation of a computation can and must be precisely defined, not dependent on an arbitrary choice of 'graining'.

While decoherence does not suffice for a derivation of the Born Rule, it does help explain why classical behavior and memories emerge. Without it, classical computations would be difficult or impossible to implement at all.

- Envariance

[Zurek] has presented an interesting symmetry-based approach to derivation of the Born Rule; unfortunately, it relies on assumptions that are not really justified. For a state like

($|A\rangle|1\rangle + |B\rangle|2\rangle) / 2^{1/2}$, the approach assumes that the probability of "A" must equal the probability of "1". This is *not* necessarily true. Perhaps the state "1" has no observers in it at all, in which case assigning *any* probability to it is meaningless. It is true that an external observer who measures the state and finds evidence of "A" would also find evidence of "1", but that tells us nothing about *internal* observers to the states we are considering. His assumption is closely related to an assumption of non-contextuality of probabilities, and results in his conclusion that equal amplitude (entangled) states have equal probabilities. The GBR with variable L-factors is clearly incompatible with this, but that does not mean it is disproven, only that it is incompatible with his assumptions.

An assumption similar to locality is then used to 'derive' the Born Rule by transforming a general entangled state to a situation with equal amplitude states, which by conservation of amplitude-squared must be in the right ratio so that counting them gives the Born Rule for the original state. The objections to relying on assumptions of locality of probabilities and noncontextuality of probabilities apply.

- Decision theory

The decision theory based approach founded by Deutsch does take into account the fact that 'probabilities' in the MWI are not abstract physical facts, but only make sense to talk about in the context of thinkers, in this case decision-makers.

[Greaves] recast decision theory in terms of caring about future branches. In her terminology, the assumptions about probability used in this paper are those based on "the Reflection argument", which she acknowledges also works. Either of these approaches seems to work as a way of understanding the nature of probability in the MWI, modulo the issues of the nature of observers and the Born Rule.

Unfortunately, while the use of decision theory might help those who have a certain type of philosophical objection to the standard way of thinking about probabilities in the MWI, it does not help when it comes down to proving that there is no quantitative *a priori* alternative to Born's Rule in the MWI.

Deutsch's approach was clarified by [Wallace]. He uses various arguments, the strongest of which being the (inadequate) Decoherence and Variable Course Graining argument, to 'establish' that branches with equal squared-amplitudes have equal probabilities. This assumption is called 'equivalence' and is used to 'derive' the Born Rule by assumptions that are equivalent to (or stronger than) non-contextuality of probabilities. All of this could be done without the language of decision theory. It amounts to an easier proof of a Gleason's-like Theorem under additional assumptions. [Gill]

- Graham's branch counting

Graham claimed [DeWitt] that the actual number of fine-grained branches is proportional to the total squared amplitude of a course-grained macroscopically defined branch. This would be great if it were true, because not only would it (together with a world-counting assumption) provide the Born Rule, it would also begin (as Gleason-based approaches do not) to give some insight into *why* the Born Rule might be true; each fine-grained branch would presumably support one copy of an observer. (That assumption would still need to be explained, of course.) Unfortunately, and even aside from the lack of precise definition for fine grained branches, he failed to justify his statistical claims [Kent], which stand in contradiction to straightforward counting of outcomes. Still, this idea founded the "count something" school of thought on the Born Rule, and the counting of implementations of computations falls into that tradition.

- Mangled worlds

[Hanson] created an interesting attempt to derive the Born Rule in the world-counting tradition. Due to the numerous microscopic scattering events that surely occur with unequal amplitudes for their various possible results, a distribution of fine-grained branches (leaving aside the question of precise definition of such) arises which has a log-normal dependence on amplitude, as it is a random walk in terms of multiplication of the original amplitude. Large amplitude branches are assumed to interfere with or 'mangle' smaller amplitude branches, imposing for all practical purposes a minimum amplitude cutoff. If the cutoff is in the right range of levels and is uniform for all branches, then due to the mathematical form of the log-normal function, the number of branches above the cutoff is proportional to the square of the original amplitude.

Unfortunately, this Mangled Worlds picture relies on many highly dubious assumptions, such as the uniformity of the 'mangling' cutoff. Nonetheless it brings to light some interesting facts, such as the tendency towards a log-normal distribution of "worlds". As it has some relationship to one possible approach to attempt to count implementations of computations for the MWI, it is discussed in more detail later.

**The Implementation Problem:**

While computationalism is a popular philosophical belief, it has been criticized. One type of criticism notes that some systems that obviously could compute seem intuitively like they could not be conscious. The most famous example is Searle's Chinese Room argument [Searle]. The intuition that some people have that these systems could not be conscious is *not* shared by computationalists, and these arguments show no internal contradiction within computationalism. This kind of criticism is not a serious problem for computationalism; some degree of counter-intuitiveness is to be expected from the implications of any theory, as physicists have long become accustomed to.

A much more serious criticism is that it seems like almost any system (such as a rock) could be considered to perform any given computation by choosing some complicated mapping between physical states and computational states. If so, then computation is not a meaningful characteristic of a physical system and can not be responsible for the conscious observations made by that system.

If one uses a naïve definition of implementation, this is indeed found to be the case [Chalmers 2]. This leads to the Implementation Problem of how to plausibly define which computations a system performs so that the wrong kinds of systems do not implement a given computation, while the correct systems do.

It is also necessary to consider how to properly 'count' the implementations of each computation, for a system that implements many different computations, so that observers can compare what they observe with predicted "probabilities" that are proportional to the number of such implementations, or to some generalization of that if there are infinitely many. This is called the Measure Problem.

In this paper, solutions to the problems of computationalism are proposed, consisting of an implementation criterion and proposals for counting implementations. These criteria are then applied to quantum physics to produce a computationalist version of the MWI, called the Many Computations Interpretation (MCI) of quantum mechanics.

If the Born rule could be derived from computationalism plus the Shrodinger equation, it would provide a strong argument in favor of the MCI. If the computationalist result definitively contradicted the Born Rule in a way that has already been ruled out experimentally, than it would show that either computationalism must be false, and/or the Schrödinger equation alone must not correctly model the physical reality. Unfortunately, it is currently *unclear* what probabilities computationalism predicts when applied to QM, due to philosophical uncertainty especially regarding how to 'count' implementations.

**Defining Implementation: Combinatorial Structured State Automatons:**

An implementation criterion will be given for a formal system consisting of a structured set of variables (also called substates), and a transition rule. Such a system will be called a Combinatorial Structured State Automaton (CSSA) and is a generalization of the Combinatorial State Automaton suggested by Chalmers [2]. This framework allows for both analog and digital computations. Quantum computations are among the analog computations.

The substate labeling used is considered part of the definition of a particular CSSA. The variables may form a vector of the appropriate dimension, or may have additional structure; for example, they may be arranged into a rectangular matrix. Substates may have no particular structural relationship to each other, in which case it will still be convenient to list them as though they form a vector, but the lack of structure will be noted. It is also possible for more than one number to be considered part of a single *composite substate* (for example, Re($\Psi$) and Im($\Psi$) may form such a composite).

The transition rule specifies what the successor state will be, given the current state of the CSSA. A CSSA may also take *input*, which means that some of the substates of the successor state are not specified by the transition rule. The transition rule may be stated in differential form (d/dt) for an analog computation.

A particular *computation* is specified not only by a CSSA, but also by the particular initial state of that CSSA and by the number of time steps that it undergoes. For example, a Turing Machine could be described as a CSSA, but it will perform different computations depending on the initial states of the tape and the head. A given computation is called a *run* of the CSSA which (along with the initial state and time steps) describes it.

A computation can be implemented by a physical system which shares appropriate features with it, or (in an analogous way) by another computation. If a physical system implements computation A, and computation A implements computation B, then the physical system must also implement computation B. In such a case, computation B is called a *simulation*, and the CSSA associated with computation A is called a *virtual machine*.

In general, a given physical system implements many different computations (e.g. the Earth implements multiple human minds), and may have many implementations of the same computation. Often a wide variety of physical systems would be able to implement the same computation (which is called *multiple realizability*).

In order for a physical system P to implement a given computation C, where C is a run of the CSSA C':

1) There must exist a mapping M between the possible states of the physical system and the formal states of the CSSA, C', which may depend on time.

2) There must exist physical laws which require that the physical system would undergo transitions which <u>would</u> cause the transition rule for C' to hold true for the states to which it is mapped, for <u>any</u> possible initial state of the physical system that falls within the mapping.

3) The physical system must (initially) be in a physical state that is mapped to the initial state of C.

4) The physical system must be correctly mathematically describable as if it were some CSSA, P', with an appropriately structured labeling of states.

5) The label structures used to describe P' and C' must allow the mapping M to be *valid* as will be described below. The key to preventing the wrong physical systems from 'implementing' a computation is to impose the correct limitations on valid mappings.

Requirement #4 might seem to conflict with relativistic invariance, because the description of a physical system as a CSSA would depend on the chosen reference frame. There may be one 'real' reference frame. However, computations of interest for the MCI should be implemented with any choice of reference frame. A change of reference frame might change the mapping, but the implementation should still work. Or if the number of implementations (see below) of each relevant computation is reference-frame-independent, as seems likely (and as the Born Rule, if derived, would imply), it would be a moot issue.

Requirement #4 does, however, rule out arbitrary labeling of physical states; in particular, one is not free to choose any desired quantum mechanical basis. (Otherwise, for a finite quantum system, one could always choose the energy basis and label degenerate states by an arbitrary index. Such a description does not provide much real information about the system, as it will simply be a group of states that oscillate sinusoidally in time.)

The physical system must be thought of as being intrinsically described in one particular basis. For purposes of applying the MCI to QM this will be assumed to be the position basis, and for quantum field theory, it will be assumed to be the basis of field configurations which are functions of position. For classical mechanics, it will be assumed to be the positions and velocities of the particles.

The restrictions on valid mappings are as follows:

Independence:

Sub-states must be independent. Criteria for Sub-State Independence (SSI) are given below.

If knowledge of the variables that determine the other substates does not reveal the value that a given substate has (at that same time step), that is *sufficient* for those substates to be independent.

If different substates depend on different physical variables, that usually suffices. However, Chalmers discovered an (exponentially growing) false implementation in which the variables record information that would reveal the values of the other states [Chalmers 2], even though different substates depend on different physical variables. This false implementation is ruled out by the above criterion.

Inheritance:

However, it is sensible to allow cases in which formal substates share physical variables that they depend on. In these cases, independence is instead established with the help of "inheritance" of labels. Physical label indices are inherited when a substate depends on the values of that index, associating the substate with that index. Permutation of the physical indices that are associated with one independent substate should not change the value of another substate.

The necessary and sufficient conditions for SSI are that at knowledge - modulo permutation of the physical labelling indices that are inherited by the given substate - of the variables that determine the other substates does not reveal the value that the given substate has (at that same time step) but does determine the others.

For example, if each independent bit $B_i$ is based on a pair of physical states $(a_i, b_i)$, inheritance is established by the facts that arbitray permutation of the a-and-b values that determine the bit does not affect the values of the other bits and does not permit knowledge of the given bit.

In the examples below, ➔ means "maps to". Most of the examples, for simplicity, are for one computation implementing another one. To modify the above implementation criterion for this, treat the CSSA states for the underlying computation as though they were physical states and the transition rules for it as though they were physical laws.

Example #1:

Bits on a 2-d grid plus time ➔ 2 integers as functions of time

b(i,j,t) ➔ i(t), j(t)

Each bit takes on the value 0 or 1. If the number of nonzero bits is not equal to 1, this mapping will be undefined. If only one bit is nonzero at a given time, then let

$i(t) = \text{sum}_{ij}\ i\ b(i,j,t), \quad j(t) = \text{sum}_{ij}\ j\ b(i,j,t)$

i(t) "inherits" the label from the i-index of the nonzero bit, and j(t) from the j-index

Knowing the physical state of the 2-d grid modulo a permutation of the j-index reveals the value of i(t), but not j(t), and vice versa. Thus they are independent, even though they both depend on all parts of the grid.

In this example, an underlying system which consists of a function on a 2-dimensional grid was reduced to a system of two variables, which together define a single point in a 2-d grid space. This can be summarized as "on ➔ in".

Note that this criterion depends on the way that the labels of the substates of the underlying system are structured. If the same number of bits were labeled by a single index plus time: b(k,t), then a similar mapping to integers i(t) and j(t) would not be valid. While that may seem odd at first, it is a generalization of rejecting a mapping from a single physical variable k(t) to two integers i(t) and j(t), which is very much what the CSSA concept is designed to do.

Note: If the pair (i(t), j(t)) forms a single composite substate, then independence is not needed and the mapping b(k,t) ➔ (i(t), j(t)) is allowed. However, because the two numbers are now considered part of the same substate, this is not the same computation as if they were considered independent. The mapping (i(t), j(t)) ➔ (separate) i(t), j(t) is not allowed (single substate to two substates).

Example #2:

Quantum harmonic oscillator ➔ classical harmonic oscillator

$\langle\Psi|x|\Psi\rangle$ ➔ X, dX/dt ➔ V

In this case the transition rule is dX/dt = V, and dV/dt = - (k/m) X. These equations will be true for a quantum harmonic oscillator.

Both X and V depend on the same physical variables, so they are not independent. The pair (X,V) must form a composite substate.

Any labeling structure of the implemented CSSA substates must reflect structure of the underlying physical (or underlying CSSA) substates. For example, it is not permitted to re-label states without proper inheritance of the label structure. One can not take the substates b(k,t) and re-label them in terms of new parameters i,j in order to 'implement' the system b(i,j,t). If this re-labeling were allowed the simulation b(i,j,t) could then implement i(t) and j(t); this would violate the assertion that the underlying system b(k,t) can not do so.

Consider a long chain of bits, b(k,t), only one of which is nonzero at a given time. At each time step, the value of the nonzero bit changes to 0, while the value of the next bit along the chain changes to 1. If arbitrary re-labeling of the bits were allowed, this could implement the system $b(B_1,B_2,…,B_N,t)$, which is a singly-occupied N-dimensional grid (SONG), which could in turn implement the system $B_1(t)…B_N(t)$, which is a set of N digital variables and which could have any desired digital dynamics. This is clearly a false implementation.

Example #3:

2 quantum harmonic oscillators (and a dial) ➔ 2 classical harmonic oscillators (in dial worlds)

This is a mapping from the physical system $\Psi(x_1,x_2,d,t)$ to classical variables $(X_1,V_1),(X_2,V_2)$. The dial has no dynamics (all orientations have the same energy) and does not interact with the oscillators, but may be entangled with them.

For some fixed value of the dial d:

$\int \Psi^*(x_1,x_2,d,t) \, x_1 \, \Psi(x_1,x_2,d,t) \, dx_1 \, dx_2$ ➔ $X_1$,     $dX_1/dt$ ➔ $V_1$

$\int \Psi^*(x_1,x_2,d,t) \, x_2 \, \Psi(x_1,x_2,d,t) \, dx_1 \, dx_2$ ➔ $X_2$,     $dX_2/dt$ ➔ $V_2$

Note that all values of $x_1$ and $x_2$ are integrated over to calculate the above brackets. $X_1$ is independent from $X_2$, since they inherit the 2-d labelling structure that had been associated with $x_1$ and $x_2$: A permutation of the $x_2$ arguments of the function $\Psi$ will leave the value of $X_1$ unchanged, and likewise for $x_1$ and $X_2$.

Suppose that the system is entangled as follows:

$$\Psi(x_1,x_2,d,0) = f_1(x_1,x_2) g_1(d) + f_2(x_1,x_2) g_2(d)$$

where $g_1$ and $g_2$ do not overlap. Fixing a value for d is like choosing one of the two "worlds" in which the classical variables like $X_1$ take on world-specific values.

Similarly, a mapping from two separate physical switches 'A' and 'B' to formal bits 'a' and 'b' preserves the independence of those bits even when the physical system is described quantum mechanically as a function on classical configuration space, and all of this can apply for some given relative state.

Example #4:

Function on a bit field on a 2-d grid ➔ a bit field on a 2-d grid

$$g(b_{11},b_{22},\ldots,b_{1n},b_{21}, b_{22},\ldots b_{2n}, \ldots, b_{m1},b_{m2},\ldots,b_{mn},t) \rightarrow b(i,j,t)$$

where g is a SONG (with N = m n), each $b_{ab}$ takes values 0 and 1,
and $b(i,j,t) = \text{sum}_{b'_{11}..b'_{mn}} b'_{ij} g$

$b(i,j,t)$ inherits its structure as a function on a 2-d grid from the structure of the labeling of the arguments of g, as g is a function on the "bit field configurations" on the 2-d grid.

Transference:

It should be possible for a Turing machine (which is largely on a 1-d tape) to simulate (and thereby implement) a computation on a high-dimensional space. This can be accomplished by allowing "transference" of labeling, by which a function of a variable can be like a function on a space; "of ➔ on".

Example #5:

Digital function on 1-d grid (+ time) ➔ digital function on 2-d grid (+ time)

$f(i,t)$ ➔ $g(x,y,t)$

Let x range from 1 to $X_{max}$.

Suppose the relations $f(1,t) = x$ and $f(2,t) = y$ must hold whenever $g(x,y,t)$ would be computed in turn for each value of x and y. The value of g is then stored in the variable $f(2+x +(X_{max}-1)*y,t)$ ; the x and y values transfer to the labels for $g(x,y,t)$.

Example #6:

Unstructured digital variables (+ time) ➔ digital function on 2-d grid (+ time)

Much like in the previous example, the key is that some of the bits will take on values that determine structure that is then transferred to the new function. For example, say that the unstructured bits, if ordered in one particular way (though this ordering is not intrinsic to their description) take on values that are identical to our digital function on a 1-d grid in the previous example, f(i,t). This system does <u>not</u> implement f(i,t) because it lacks the intrinsic structure, but it <u>does</u> implement g(x,y,t) because it satisfies the conditions needed for "of ➔ on" transfer. Ordinary computing machines as well as human brains are made of components with no intrinsic label structure, but they *can* implement simulations of structured systems.

<u>Notes</u>:

Counterfactuals:

In order for counterfactuals to be meaningful, the fundamental laws of physics can not be just summaries of relationships between states. The laws themselves must actually exist, just as physical states do. Likewise, the transition rule is part of the specification of a CSSA. It is assumed that the physical world could be described mathematically as a CSSA. Generalization of this could be done but is not needed for currently used models of physics; quantum gravity may require such generalization, but that remains to be seen.

A "timeless" generalization is not hard to make. Formal "time steps" would rely on mappings from different physical regions, not different times. For example, this could be done if physics were a boundary value problem instead of an initial value problem. The transition rule must hold true: If an appropriate change were made in the first region, the laws of physics must imply that there must be the corresponding difference in the next region. It must still be the case that there are many possible physical states consistent with the laws of physics, so that counterfactuals hold true. While that works formally, it seems to violate the spirit of computation, which is embodied by initial value problems. It is more plausible that physics is in the form of an initial value problem. The possible "frozen formalism" of quantum gravity, its problems, and ways that physics could avoid it will be discussed later.

The digital clock:

Suppose that an N-bit clock counts in base 2, turning back to zero when its limit is reached. If any starting state of the clock is chosen, it is certain that at some point, the clock will enter any other given state. This may seem to make it satisfy all possible transition rules. Does it implement all N-bit computations?

It certainly does not, although there are some options regarding what rules should be used to prevent it from doing so. The most straightforward is that the transition rule must specify the next mapped-to formal state in time, while the clock goes through intermediate states that the mapping would recognize.

However, mappings tend to be defined only at discrete times (or narrow ranges of times) corresponding to time steps. The solution I will use is based on that case, and it is simply that the clock fails to implement the counterfactuals. If a pair of times $t_0$ and $t_1$ are chosen such that the state at $t_0$ is A and the state at $t_1$ is B, then the activity of the clock looks like "A then B". However, if the clock were placed in state C at $t_0$, it will be in some state C+k at time $t_1$. It can not implement an arbitrary computation which might require the transition "C then D". There is no choice of times $t_0$ and $t_1$ that will make the clock satisfy the transition rules for all possible initial states of an arbitrary computation.

It is not a problem if mappings are defined over ranges of times, so long as the mapping specifies unambiguously what the formal state will be at each formal time step. It is never permissible for a mapping to assign more than one formal state to the computation at the same formal time step.

Suppose the speed of a computer depends on the state of that computer. In this case, the mapping might well specify that the next mapped-to formal state is the one that applies to the next time step. However this would be specified by a particular mapping, not necessarily being a general rule.

Could a mapping specify that the next formal state should be evaluated at a time t(S) that depends on the current state S? No; a mapping for one time step can not use knowledge about the formal state of the previous one, and the digital clock shows why.

The derail-able computer:

[Maudlin] has argued that consciousness supervenes only on physical activity, while computation requires counterfactuals and therefore depends more on physical structure. He gives an example of a computer that operates normally for one input, but is required to call on different machinery for any other input. He claims that if the computer's parts are disconnected so that the secondary machinery is not available the system should still be conscious if running on its normal input, and concludes that only actual activity matters, not counterfactuals.

However, consider a 2-d cellular automaton (CA) with 'nearest-neighbor' interaction, but which has no visible grid structure that defines which nodes are neighbors. (Imagine tangles of wire connecting a bunch of electronic switches with each acting as a node, piled randomly in a chaotic heap, with plenty of extra connecting wires which send unrelated signals.) Looking only at the activity of the switches reveals little about the cellular automaton, since there is no grid to look at. However, if one takes into account the actual causal relationships between switches, which are correct and reflect some 2-d grid, one could reconstruct the description of the system as that cellular automaton. It is true that one could do so based on activity in the wires, but tracing causation is the *reason why* that would work. Now consider switches which act randomly, but for some time happen to act the same as our randomly tangled up CA; surely these switches do not implement that CA, which has no relation to their arrangement.

The 'activity hypothesis' must be rejected; however, if a computer is built that 'derails' for the wrong input, that does not mean the computer does not implement any computations. It is true that it will not implement the same CSSA as it would if it did not suffer from the derailment issue, but it will still implement some CSSA which is related to the normal one. This new CSSA may be sufficient to give rise to consciousness. If, however, the system simply 'replays a movie' – that is, displays the right activity but does not check at every step for derailment - then it does not implement any interesting CSSAs, and it can not give rise to consciousness.

The huge look-up table (HLUT):

A computer could be built which transitions to a successor state by looking up the current state on a huge table and finding the corresponding successor. If done in this way, skipping no steps, this computer would be a legitimate implementation of a CSSA. The mapping would have to require that the HLUT is in the correct physical state. However, if steps were skipped so that the HLUT's results are those corresponding to many time steps later, then the system would not implement the original CSSA.

The environment:

The environment of a computer must be 'friendly'. In other words, if there are incoming bullets, an implementation of a complex computation is liable to get ruined. As a result, in order to describe transitions that are 100% reliable, more description of the 'environment' of a computer is needed as part of the mapping than one might think. This can be done, of course, but it is common to neglect it in describing examples of computers.

**Counting Fundamental Conscious Implementations; Effective Probabilities:**

The "measure" of a type of conscious observation is defined as the quantity of conscious observations that have the given experience. The "probability" that an observer will be a given computation (given some reference class characteristics of the observer) is equal to the measure of that computation divided by the total measure of such computations in the observer's reference class. [Bostrom]

Such a "probability" represents the best guess that an observer could make, given the nature of the physical system that gives rise to him. Taking into account additional information could be used to find conditional probabilities. In the simplest case, his ignorance of which computation he is performing is due to not knowing what input he is about to receive. This can be considered merely a Bayesian subjective probability, since there is no actual randomness involved, and using this procedure allows the maximum number of observers to guess correctly about any given conditional.

Note that the total measure of all conscious computations is not generally conserved as a function of time. Even in a situation where the number of observers would classically remain constant (that is, observers remain conscious and there are no births, deaths, or macroscopic 'clonings'), it should not be assumed *a priori* that the measure of observers remains constant.

The implementation of a *conscious computation* gives rise to a conscious experience. The basic idea that would permit predictions to be drawn from a computationalist model is that the measure of each type of conscious computation is proportional to its number of implementations. This idea requires a few refinements before it can be applied.

One computation may simulate some other computation and give rise to conscious experience only because it does so. In this case it would be unjustified double counting to allow the implementations of both computations to contribute to the measure. This problem is easily avoided by only considering computations which give rise to consciousness in a way that is not due merely to simulation of some other conscious computation.

Such a computation is a *fundamental conscious computation* (FCC). It is possible, theoretically, for one FCC to simulate another (such as by a person visualizing the extended operation of a Turing machine implementing another FCC, although no human is intelligent enough to do so), in which case both FCCs must be dealt with separately.

If the number of implementations is infinite, it is not a well defined procedure to take a ratio of the numbers to determine probabilities. However, this can be circumvented by either a symmetry argument or imposing a "natural" regularization on the set of implementation mappings that would make the numbers finite, and then taking the limit as the numbers are allowed to approach infinity. For example, for an infinite universe this might be done by considering mappings limited to a succession of larger spatial boxes, and allowing the box size to approach infinity.

A choice of regularization can affect the limits that are obtained, so one must argue that the choice is natural or "inherited" from the true mathematical description of the physical system. The symmetries of the system might come into play. This is a critical philosophical issue, because the derivation of the Born Rule may rely on invoking symmetries if that is the correct way to "regularize" the ratios of infinitely many implementations.

There is another issue that must be dealt with, that of how the size or flexibility of a system affects the number of computations it implements. This is called the Problem of Size.

Suppose that switches in a system can take values from $-n$ to $+n$. All switches are set to some value, $-s$ or $+s$. The system then acts as though each switch is a bit, ($-s$ as $-1$, $+s$ as $+1$), and the output switches take on the value $+s$ or $-s$ as appropriate for bit values $+1$ or $-1$. How many implementations of the bit-based computation does this system perform?

One could choose a mapping that only includes some possible values of the switch magnitude, such as "1,3,6,11,15,or n". This mapping is undefined if the switch magnitude is not in the list. There are many such possible lists, in fact $2^{n-1}$ possible lists that will include any given value of s. Any one of these mappings does lead to a legitimate implementation.

It seems obvious that the above system performs the given computation only once, and not $2^{n-1}$ times over. And certainly it is to be hoped that minor details of a conscious being's construction does not have such a large effect on its measure of consiousness; otherwise a 'large switch brain' would be equivalent to a large population, and such a being could rightfully demand voting rights and power in line with its equivalent population.

The problem with counting each of the above mappings seperately is that they are not independent implementations; for example, it is not possible for them to start off in different formal states from each other. A Measure Counting Implementation Independence (MCII) criterion is needed.

There can be no subjectivity involved in choosing mappings; therefore one must count all possible mappings that satisfy the independence requirements, subject to the rule that a correct set of independent mappings should *maximize* the number of independent implementations.

The measure of a FCC is equal to the maximal number of its independent implementations. This is well and good for classical computers, in which the number of implementations is typically proportional to the actual number of machines (or brains) running that program, and does not depend on the type of hardware.

However, it may imply that the measure of a quantum mechanical "branch" does not depend on the amplitude – in contradiction with the experimentally observed Born's Rule. (Note that "does not depend on amplitude" does not mean the measures must all be equal; there could be other factors.) These issues can be explored for specific choices of the MCII criterion.

MCII #0: No restrictions. All valid implementations are to be counted.

This possibility is included for completeness. As noted above, if there are n possible states of a physical substate, there are $2^{n-1}$ combinations of the states that include any given state. This will not give the Born Rule for the quantum case, because the number of implementations would be independent of amplitude.

A natural choice for MCII might be to use the same criterea as for Sub-State Independence:

MCII #1: Implementations of a given computation are independent if when their mappings are simultaneously applied to the same physical system, their respective formal variables are independent, using the SSI criteria. This can be viewed as a single implementation of a large computation that contains many copies of the given computation.

However, there is no reason that a different MCII criterion could not be used, because implementation counting is not the same thing as substate distinguishing. Another possibility is the following:

MCII #2: Implementations of a given computation are independent if when their mappings are simultaneously applied to the same physical system, it would be possible (if one could choose initial conditions for the physical system) to start off each implementation in any of its formal states without restrictions due to the states of the others.

The MCII#2 critereon takes no notice of the structure of the physical states. Note that this type of critereon can not be used as the basis of sub-state independence within an implementation, because that is just the naïve implementation critereon which is vulnerable to false implementations. But a basis for implementation counting, given that each individual mapping must also satisfy all criterea for implementing a given computation such as SSI independence of its substates, it could work.

Though bearing in mind that independence criterea are of a general nature and not defined in terms of a specific model of physics, it must be noted that in a quantum mechanical context, the number of MCII#2-independent mappings if each is based on a component of the wavefunction is the same as if these components must all be orthogonal to each other. In order to actually implement a run of a computation, the components in question would also have to not be orthogonal to the actual wavefunction.

The mappings must be chosen so that the number of independent implementations is maximized. If this number would be infinite, a 'natural' regularization of the mappings must be chosen, then a limit towards infinitely many implementations is taken to find measure ratios. Note that such a regularization applies to mappings only, not to the actual physical system, which does not take notice of any mappings that are made to understand its implementations of computations. For example, for an infinite universe with a finite box regularization, one might use a density matrix to generate mappings within the finite box, but there is no need to come up with new physics to describe the goings on at the edges of the box; the actual physics holds, as the box is not a regularization of the physical system, just a limitation on the mappings being made.

An example in which MCII #1 and MCII #2 can give different results is:

Suppose $\Psi(s,x) = a\, f(+,x) + b\, g(-,x)$
Let $X+ = \int x\, |\Psi(+,x)|^2\, dx$, and $X- = \int x\, |\Psi(-,x)|^2\, dx$.

Suppose that if a and b remain constant, then both X+ and X- implement harmonic oscillators.  X+ and X- are independent in the SSI sense, because they depend on different physical variables (assuming that different spin states are not a single composite variable).

However, if there is a spatially uniform but time varying magnetic field that causes a and b to change in an unpredictable way, then with MCII#1 X+ will no longer implement a harmonic oscillator, and neither will X-.  Their sum will implement a single harmonic oscillator.

With MCII #2, a fix would be to project $\Psi$ onto a time-dependent unit vector which tracks the rotating spin, so that the original spin amplitudes apply in the new coordinate system.  This can be the basis for two independent mappings.

The case where a=1 and b=0 seems pathological in that MCII#1 now gives only one mapping that implements the harmonic oscillator.  MCII#2, by allowing projections onto other spin directions, removes that problem.  However, in practice no spin would ever be perfectly aligned with the z-axis; it is a zero-probability event.  In the particular case of spin-1/2 particles, in order to enforce rotational invariance with MCII#1, one could assume that $(\Psi(+,x), \Psi(-,x))$ form a single composite substate, so that there is always only one implementation of the harmonic oscillator in the above example after all.

Activity in distant regions of space would seem to be something that should have no effect on the number of implementations within a spacially limited computer.  If Born's Rule is derived then it would imply that it does have no effect, but what do the implementation independence critera imply about it?

With MCII#1, occupation of previously zero-valued regions of configuration space increases the number of implementations, just as it did the the spin-1/2 harmonic oscillator example with a=1 and b=0 changing to a state with b > 0.  This effect would allow distant events to affect the measure of our computer, and in principle it could be correllated with the outcome of a measurement the computer performed, thus negating the Born Rule.

However, every region of configuration space is bound to have some amplitude in it, however small.  In the spin-1/2 case the two independent harmonic oscillators were ruined, but if the distant action is a macroscopic object moving and our computation of interest is digital, one can choose time dependent mappings so as to approximately preserve the number of implementations, which must always be maximized.

Also, note that this nonlocality may be a problem not so much with MCII#1 as with the Schrödinger picture ontology, which is defined in a given reference frame.  If the local Heisenberg picture is used for the MWI, an explicitly local formulation would result. [Rubin] However, the Born Rule is the real issue, and it seems more likely that a sort of generalized 'APP'-like rule would result, with measure proportional to phase space volume rather than amplitude-squared. The Heisenberg picture does still require detailed investigation.

It is possible that the true nature of the physical system is neither the Heisenberg picture nor the Shrodinger picture, but some other mathematical model, and that this true model implies the Born Rule from MCII#1. But if the model strays too far from the standard formulations that approach risks being considered a many-worlds hidden variables model; certainly a fallback position to consider, but not the possibility that an Everett fan would hope for. It is also possible for interference from other branches of the wavefunction to affect the number of implementations, perhaps giving an approximate Born Rule; inspired by the Mangled Worlds idea, this possibility will be explored below.

With MCII#2, once again as in the spin-1/2 example, it is possible to project onto time-dependent orthogonal relative states that describe the distant activity, so as to preserve the number of implementations in the local computer.

The number of orthogonal states needed to describe a given system within a fixed volume may be finite, as shown by the maximal entropy of black holes. One would think that within a finite-box-limited regularization, the maximum number of such states should be used. In this case, except for the issues of "noise" as will be discussed below, the number of implementations appears to be independent of amplitude (see the schematic implementation that will be presented later for a more detailed analysis). This assumption relies on poorly understood physics, however, as well as the fixed volume assumption. Indeed, if a black hole is moving at high speed, its event horizon would be Lorentz contracted and occupy a smaller volume, so presumably the simple assumption above is not the whole story.

A more general objection is that applying the finite-box regularization in that way is not appropriate; the actual physical system is never changed by a mapping regularization, and if the universe is infinite or potentially infinite with the right initial conditions, so is the number of physical states. A density matrix might be used to base mappings on, but *not* simply a spacially truncated physical system, which would be a different physical system and not merely a mapping limitation. A better finite box regularization might be applied only to the parts of the mappings that are based in the brains, leaving the distant parts of the wavefunction to be regularized some other way. Doing it this way would make the order of the two regularizations unimportant. That seems an advantage; the finite box regularization is then only needed because there could be infinitely many distant planets with the same brain types as ours.

It may be that the wavefunction of the universe is translationally invariant, with correlations providing the appearance of breaking the symmetry. For example, $\Psi(x_1,x_2,t) = f(x_1-x_2, t)$ would be such a function that has overall translational invariance. A finite box regularization would not seem natural for such a function.

If the number of states (and thus MCII#2-independent-implementations) is infinite, some regularization or symmetry argument must be used to find measure ratios. There is one possibility that may have some intuitive appeal, and would yield the Born Rule:

Equal Norm Regularization Conjecture (ENRC):
When an implementation mapping regularization is used that for any finite stage of the regularization determines the maximum number of orthogonal components of vectors in an infinite-dimensional space, that regularization should assign an equal norm ε to each of the orthogonal components. When comparing vectors and the number is not an integer, round off (up or down won't matter in the infinite-number-of-mappings limit).

This is equivalent to assuming that at each stage of the regularization there is a minimum-norm difference ε between distinguishable vectors. [Buniy et al] suggested a similar thing as a possible modification to physics to yield the Born Rule, but here it is suggested not as a law of physics with an actual minimum norm, but instead as a natural type of mapping regularization, with a minimum norm that vanishes as infinite-number-of-mappings limit is taken for the regularization.

It may not seem plausible that a regularization could be the source of probabilities. That seems much more plausible in the case of a finite-box regularization for an infinite universe, however, and that establishes a precedent. The real questions are whether the ENRC is relevant to the physical situation and is plausible if so. By placing no emphasis on the physical state structure that the vectors are functions of, the ENRC does seem to fit the spirit of MCII#2 as opposed to MCII#1.

Consider the inverse problem: Given a finite but ever increasing set of orthogonal vector components, what is the most natural choice for the squared norm of the resulting vector? Using a squared norm that is proportional to the number of components, as it would be if they each have equal norms (because of the Pythagorean Theorem), seems reasonable. If so, the ENRC may seem plausible as well. In any case, the ENRC does provide a definite regularization, and without a regularization the probabilities can not be evaluated.

A more detailed look at quantum mechanical implementations of computations will be given later, but the argument that the ENRC would yield the Born Rule (ABR) (with certain assumptions) is roughly based on the following. Consider the vector $|\Psi(t)\rangle = a |A(t)\rangle + b |B(t)\rangle + \ldots$ in an infinite-dimensional configuration space, where the $|A\rangle$ and $|B\rangle$ terms (given the absence of any interfering terms) implement FCCs. At a finite stage of the regularization, with the same ε, the $|A\rangle$ term has roughly $|a|^2/\varepsilon^2$ orthogonal components, while the $|B\rangle$ term has roughly $|b|^2/\varepsilon^2$ of them. MCII#2 relates these orthogonal components to the independent implementations that these terms could perform.

One might object that the ENRC was chosen precisely because it is equivalent to the Born Rule, with little independent motivation. The objection is not without merit. What the MCII#2 + ENRC may prove is not that the MCI *must imply* the Born Rule, but that it *could imply* the Born Rule and *need not* imply some wrong rule, which was not at all obvious given the troubles with MCII#1. That alone is important because it would take the Born Rule off the table as a show-stopper for the MCI, which is otherwise appealing on grounds of simplicity and compatibility of a theory of consciousness with reductive materialism.

While that may be acceptable, a true derivation that leaves no doubt of its *a priori* nature would certainly be more desirable. For that reason, other attempts to derive the Born Rule within the MCI are investigated below.

Another issue to consider is that of constraints. For example variables v1, v2, and v3 may be constrained so that v1 + v2 + v3 = 0. A mapping that relies only on v1 is independent of a mapping that relies only on v2, but what about adding a mapping that relies on v3 to the set? If *binary independence* suffices, then all three may be simultaneously included in the implementation count; by default, however, it will be assumed that all mappings must be independent of all others simultaneously, as opposed to only comparing one pair of mappings at a time.

There is however another issue to consider, and it could lead to further modification of the choice for MCII.

Consider a system which undergoes the following sequence of states:

time→ ---A---A'---B---B'---

Assume that mapping A' to an A-like state is no problem, and likewise B' can be mapped to a B-like state. The states may have complicated substate structure (i.e. bitstrings) so that mapping A to B would trivialize the computation by ignoring substate independence.

How many independent implementations of the basic computation (A transitions to B) should I think this system performed in the above sequence?

The obvious answer would seem to be one: (A transitions to B) and (A' transitions to B') don't seem independent. I can't change just A to some other state (say, C) without also having A' change, so MCII#2 rules it out, and MCII#1 certainly would since knowing A reveals what A' will be.

Now consider the following sequence:

time→ ---A---A'---B---B'---A---A'---B---B'---

Because there are two cycles, I now want to say that there are two implementations of the computation (A transitions to B). But the above arguments now seem to apply as well, so there is still just one implementation. However, there is now the three-step computation (A then B then A then B).

Which is fine, but there seems to be a sense in which this three-step computation simulates two copies of (A then B). If (A then B) is somehow a FCC, I'd expect it to have the measure associated with two implementations, not just one. So I'd need another rule to get that information, which is not too hard to find.

But it might be that unavoidable cycles really don't count as additional implementations; nothing new is really figured out by repetition, so it could be plausible, though it seems strange. One advantage of such a view is that if all mathematically possible universes really exist [Tegmark] it could help explain why simple cyclical universes don't dominate the measure distribution of FCC-implementing systems, since multiple cycles wouldn't count. However, the all-universes hypothesis (AUH) would still have the problem of selecting a measure distribution.

Alternatively I could bite a bullet and allow that in a sequence like the above there are actually four implementations of (A then B). Even that requires some restrictions, of course; it seems obvious at least that the measure should not grow exponentially with the cyclical sequence length, but rather linearly.

Another transition structure to consider is branching such as

```
A----- B         or      A---B''--- B
  \--- B'                    \---- B'
```

This could be an irreversible computation; if B is changed but B' is not, there may be no initial state compatible with that. Known physics is reversible, but irreversible computations can certainly be implemented by reversible systems (and most classical computations are irreversible), so that is no reason to rule out such a structure.

The branched structure above shares a common initial state, so it can not represent two independent implementations of (A then B) … or can it? There could be a case in which each B-type state was figured out "independently" of the other from the same A-type starting information. Perhaps making this distinction is not possible, and such branched structures always represent multiple implementations. Pursuing this line of reasoning, an implementation independence critereon could allow multiple implementations with the same initial state but different final states.

This is a somewhat strange conjecture if it allows making copies of a final state (or allowing a final state to persist for a longer time) to generate more implementations. This is not what one would expect for independent implementations, and seems like a step back towards Maudlin's activity hypothesis. Nonetheless, if it would yield the Born Rule, it is interesting to study the possibility.

Some other critereon should be used to limit the final states. One idea that has some appeal is a restriction to different value 'bands' (of physical states or inherited values) for final states for different implementations that share the same initial state. This would allow waves with greater amplitude to have more implementations because additional 'bands' are reached by final state waves.

MCII #3: Use MCII#1 (SSI-type independence) for the initial state of a time step, but mappings with the same initial state that rely on different "physical ranges" for the final state are considered independent.

A "physical range" is set of values for the physical variables that the mapping is based on, and is not changed by permutation of those variables.

This might be used to 'derive' Born's Rule if certain conjectures hold, as follows:

The initial state of a mapping will map some set of wavefunctions to an initial formal state A. For the final state (after one time step), it can be known with certainty that if the system was in a state that mapped to A, the wavefunction must have certain properties at this later time, which makes it fall into a region of physical variables that is mapped to formal state B. (Counterfactuals must hold of course, so similarly for other formal states such as (C then D).)

It is conjectured that the mappings can be chosen so that a finite region on the ($Re(\Psi)$, $Im(\Psi)$) plane is guaranteed to be occupied during the final time step when one integrates over space at a fixed time, if A was the initial formal state. In other words, the final state acts as a 'branched structure' because it is spread out on the space of physical states, and final state mappings are no longer required to satisfy the stronger independence criterea of MCII#1 or #2.

There are thus an infinite number of choices for final state mappings (corresponding to any point in that region), so (as was done for the case of an infinite universe) a regularization must be invoked to find probabilities, using some natural measure or symmetry of the mathematical description of the physics. For example, a digitization of the wavefunction would work as a regularization, with points uniformly spaced on the ($Re(\Psi)$, $Im(\Psi)$) plane being a natural choice that indicates the measure is proportional to area on that plane, and thus to the amplitude$^2$ of the wavefunction for these particular mappings.

For a finite box regularization with a finite number of quantum states to work with (due to the Quantum Gravity (QG) black hole entropy bound), at any one time there would only be a finite number of 'final state' variables so that even MCII#3 would yield probabilities independent of amplitude if the final state is restricted to a fixed time. However, if ranges of time are allowed, then once again it yields the Born Rule since the number of points on the ($Re(\Psi)$, $Im(\Psi)$) plane passed through in finite time is again proportional to amplitude$^2$. Alternatively, even if the final state mapping is restricted to a single time, and the number of quantum states is finite, the Born Rule can be obtained by a simple modification of the physics: addition of interpolating segments (e.g. linear segments) between quantum state amplitudes to produce a continuous function.

| | |
|---|---|
| / | Bands of amplitude: |
| /← | Interpolating segments could make Ψ a continuous function even with a finite (QG) |
| / | number of quantum states to work with. This would be a simple 'hidden variables' |
| / | modification of physics, where the 'hidden variables' occupy all branches. |

Additional mappings could be made using other parts of configuration space, but that possibility is amplitude-independent. For different branches of the same total squared amplitude, it may be that there are a different number of configuration space regions that could be used, but the fact that more are occupied would decrease the amplitude within each one. So it may be that the total volume integrated over configuration space of the appropriate area on the (Re(Ψ), Im(Ψ)) plane is what matters; if so that could give the Born Rule, or an approximation to it. The need for a regularization may even introduce a back door by which the symmetries of the theory of relativity should be brought in, and the total amplitude-squared is an invariant quantity.

Regardless of any philosophical distaste, it might be thought that the MCII#3 scheme would be falsified experimentally since some branches become more correlated with the environment (allowing more final state mappings) than others, but that is not necessarily so: Making additional records in quantum mechanics increases the amplitude in some regions of configuration space but decreases the number 'classical configurations' that the amplitude is spread over, as compared to leaving the environment alone and thus having a product-form wavefunction.

It must be noted that once the brain state changes again, crude records (e.g. MRI scans) that do not preserve the important brain state information and independence relationships would be of no use for trying to support B-type mappings if B is a state of a FCC.

For most of this paper, it will be conservatively assumed that MCII#3 is false, and an attempt will be made to derive the Born Rule less directly but from the MCII#1 or #2. This will be seen to have its own difficulties, so the possibility of Direct Derivation of the Born Rule by some improvement upon the above argument, or by some more clever argument that has not occurred to this author, should be considered.

**Wave Implementations and QM:**

For a wave system that implements computations, care must be taken to try to find all the mappings that meet all of the requirements for implementations and for independence. This is especially true for quantum systems because their mechanics occur in high-dimensional Hilbert spaces and are hard to visualize.

First of all, could a wave in fact implement a classical digital computation? A set of particles such as billiard balls with appropriate interactions and walls to bounce off could implement a classical digital computation, so to answer this it suffices to show that a wave system could mimic such a set of particles well enough that the implementation of the computation would work.

Wavepackets in which the wavelength is small enough that diffraction and interference will not ruin the computation should suffice for such a computer. As an example, the expectation value of the position of a wave packet (a function on space, e.g. a sound wave) could be used to define a "particle" position for a computation based on classical mechanics such as for billiard balls. If the "particle" is within specific regions, the mapping puts the computer in the appropriate formal substates.

Note that while the counterfactual relationships for a mapping have to be perfect, a computer itself does not; it is OK if the relationships would fail some time after the time steps in question occur. Note also that mappings are allowed to be time-dependent, so it is OK if the mappings for the final time steps are not quite the same as for the initial state.

The latter property of mappings is not only useful for wave packet spread; in the case of a QM system, the system may be moving at some velocity and even undergoing accelerations, so the regions of space on which a given substate depends on will be different at different time steps. This is even more important for QFT, in which one does not have the luxury of assuming that each physical switch consists of its own fixed set of distinguishable particles. Causal relationships and position-basis-inherited independence of variables must determine what mappings can be used.

If the wavelength is short enough, a wave packet can act like a billiard ball, so it can implement computations. But how many implementations does a wavetrain perform?

A wave packet is basically characterized by its amplitude, longitudinal and transverse widths, wavelengths, and environment. Of these factors, the amplitude intuitively might have no effect on the number of implementations performed by a strictly linear system.

The wider the wave packet, the more room there is for counterfactual changes in the physical state to allow different independent initial formal states to lead to distinguishable independent final states. Such changes would leave gaps in the wavetrain and would produce new wavetrains appropriate to the counterfactual states, but if the wavelength is short enough, the waves can still travel a good distance before diffraction ruins the billiard ball-like behavior.

The effective volume of the wave packet is therefore a key factor. In a very-high-dimensional space (such as the phase space for QM), almost all of the volume of an object occurs near its outer edges. The outer edges of the wave packet will therefore determine the effective volume. If the wave packet is perfectly isolated, this will depend only on the fall-off of the wave packet itself.

However, if any noise or interference from other wave packets is present in the medium, the situation may be different. In this case a higher amplitude increases the effective width of the wavetrain, beyond which the noise would overwhelm the desired implementation. For example, the wavetrain may have a Gaussian envelope, with tails eventually getting lost in the noise.

Larger volumes can be used to aggregate "signal" so as to overcome noise. The "signal" relevant for the mappings could be $<\Psi_{wp}|f>$ where $\Psi_{wp}$ is our wave packet and f is some function that projects onto the relevant phase space volume and wave travel parameters. This "signal" would be proportional to the local amplitude and to the volume V used for the mapping. Assume that the noise is uncorrelated on scales comparable to the basis for the mappings considered. The average "signal" to noise ratio ($<\Psi_{wp}|f>$ / $<noise|f>$) is then proportional to $(A\ V) / (n\ V^{1/2})$, where A is the wavetrain amplitude and n is a noise level parameter. If a minimum "SNR" is required to assure the counterfactual relationships for an implementation

$$\text{"SNR"} \sim (A\ V) / (n\ V^{1/2}) \quad \text{so} \quad V^{1/2} \sim \text{"SNR"}\ (n\ /\ A)$$

this implies the minimum volume required is proportional to $n^2 / A^2$. Since the number of implementations is inversely proportional to the volume required for each one, this gives the interesting result that the measure is proportional to $A^2 / n^2$.

In the above model, the portion of the integral of the squared amplitude within the noise-limited tails determines the number of implementations. While suggestive, any attempt towards derivation of Born's Rule must deal with many issues.

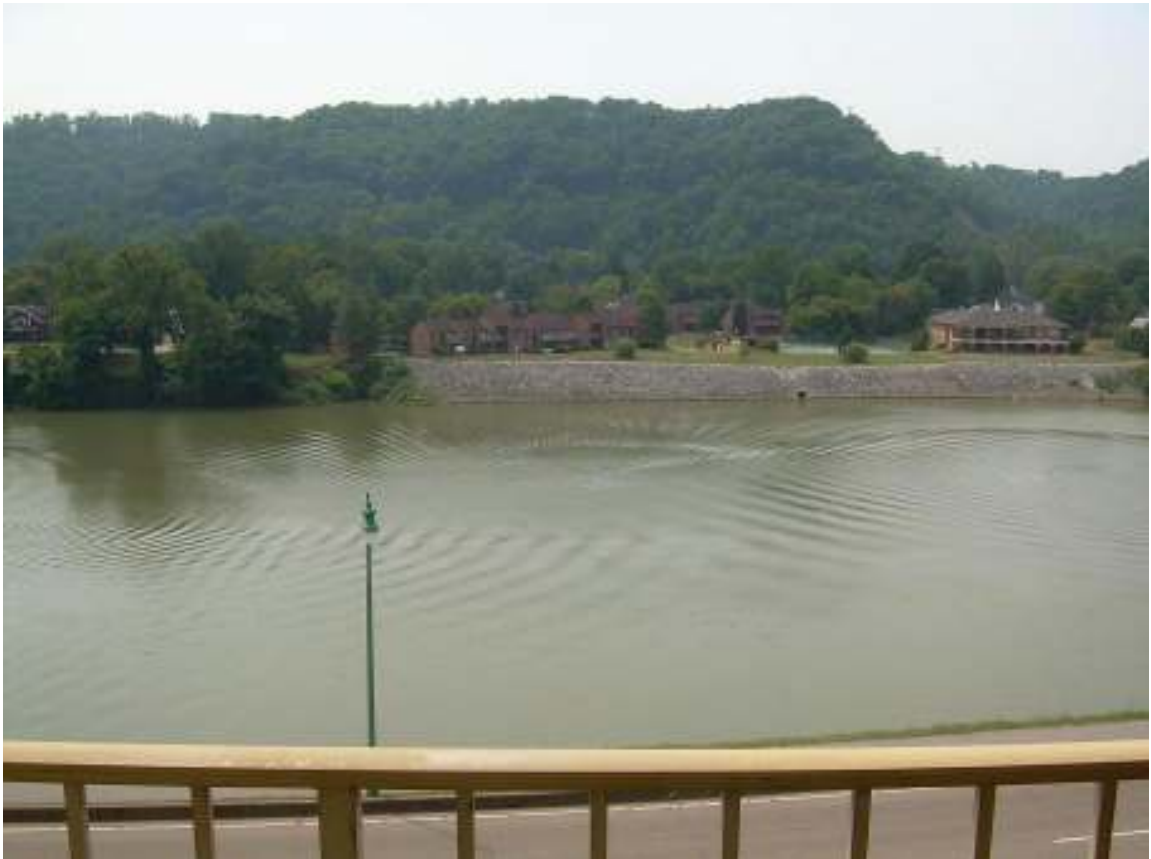

**Waves can implement particle-like trajectory-based computations in parallel along a wavefront. Diffraction that would happen if independent implementations counterfactually start in different states and 'noise' interference from other waves can limit the number of implementations with MCII#1 or #2.**

**Quantum Mechanical Implementations:**

It is important to note that, in order to implement a meaningful computation, it is not sufficient that the wavefunction be non-orthogonal to some Hilbert space vector. Assume |a> is a wavefunction that does implement some FCC, and that energy eigenstate |e> is not orthogonal to |a>. It is true that <e|a> is nonzero, but |e> can not give rise to any nontrivial implementations.

Note also that in order to satisfy the requirements for an implementation, it can not be possible for the mapping to allow the implemented CSSA to be in conflicting formal states. For example, suppose that a system in state |0> maps to a given classical bit taking the value 0, while state |1> maps to the same bit taking the value 1. One might think that the state (|0> + |1>) would then indicate that it has both values at once. But this is a classical bit, so that is not possible. One mapping that could work is as follows: For a system in state |s>, the bit takes value 0 if |<0|s>| is greater than |<1|s>|, and it takes value 1 if |<0|s>| is less than |<1|s>|. If they are equal, this mapping is undefined.

If the original state is |0>, then adding the vector |1> to that state ruins the implementation. This shows that orthogonal (but coherent) quantum states can interfere with the implementations that the other states would have performed.

In order to satisfy the counterfactual requirements for a classical computation, it is generally necessary that states of the physical system that would map to counterfactual formal states would decohere from the actual state of the physical system; otherwise the system would behave as a quantum computer, not a classical one.

The total measure of observers is not in general conserved as a function of time, and need not be conserved if a system is subject to some unitary operation. However, <u>aside from classical population differences</u>, there is good empirical reason to believe that total measure can not be a strong function of time. Suppose that it were; in particular, suppose that every time a "branching" event occurs in the universe the measure of an observer doubles, as one might think if measure is proportional to the number of branches that contain an observer. If this were true then the measure of a person living in the year 2017 would be very much greater than the measure of a similar person living in the year 2007, so that the relative probability of finding oneself in the earlier year is completely negligible. At this writing, this observer is in 2007, thus experimentally falsifying that hypothesis with very high probability. A proper derivation of Born's Rule must explain not only the branching ratios but also the approximate conservation of measure as a function of time. If the number of implementations (and thus the measure, not merely the probability of a branch) is proportional to the squared amplitude of that branch, that would explain it.

**Schematic example of quantum mechanical implementations:**

Taking the above observations into account, a slightly more detailed schematic example of how computations would be implemented by quantum systems can be given. The Shrodinger picture will be assumed.

Dirac notation will continue to be used, but it will be assumed that the wavefunction is ontologically real, not just state kets. In other words, a term like |A> represents the function on configuration space <spins,configurations|A>. Terms like |A>|1> represent functions that are products of functions in the respective subspaces.

Suppose the wavefunction at time = 0 is

$|\Psi_0> = a\ |b_{10}>|e_{10}>|s_{10}>|g_0> + b\ |b_{20}>|e_{20}>|s_{10}>|g_0> + c\ |b_{30}>|e_{30}>|s_{20}>|g_0> + d\ |b_{40}>|e_{40}>|s_{20}>|g_0> + e\ |rest(0)>$

and at time = t it is

$|\Psi_t> = a\ |b_{3t}>|e_{1t}>|s_{1t}>|g_t> + b\ |b_{4t}>|e_{2t}>|s_{1t}>|g_t> + c\ |b_{5t}>|e_{3t}>|s_{2t}>|g_t> + d\ |b_{6t}>|e_{6t}>|s_{2t}>|g_t> + e\ |rest(t)>$

where

$b_i$ = a brain state
$e_i$ = the near environment, closely entangled with the brain states
$s_i$ = a more distant environment, entangled with groups of brain states
$g$ = a far environment, not entangled with the brain states of interest
rest = other terms of the wavefunction, assumed for now not to interfere with the brain states of interest

Such a wavefunction is not completely realistic. Large, complex subspaces will be entangled, so $|b_{10}>|e_{10}>|s_{10}>|g_0>$ is an idealization. Also, there is always some small amplitude in other terms that could interfere with these, such as $|b_{3t}>|e_{1t}>|s_{2t}>|g_t>$.

For this simplified example, the one-step run "$B_1$ then $B_3$" of a computation ($B_i$ then $B_{i+2}$) is to be implemented. The computation has a detailed substate structure, which is not shown explicitly here; "$B_i$" is just a shorthand for one complicated formal state.

First, at least one mapping must be found which satisfies all of the requirements. The following mapping can be used:

Let

$P_{bi} = (|b_i><b_i|) \otimes 1_e \otimes 1_s \otimes 1_g$

$P_{Si} = 1_b \otimes 1_e \otimes (|s_i><s_i|) \otimes 1_g$

$F(i,j) = <\Psi|\ P_{bi}\ P_{Sj}\ |\Psi>$

The formal state at time = 0 is $B_i$ if $F(i,1) > (1 + \delta) \Sigma_{k \neq i} F(k,1)$

The formal state at time = t is $B_n$ if $F(i,1) > \Sigma_{k \neq i} F(k,1)$

Require also some condition for safety like $\langle \Psi | C | \Psi \rangle$ less than $F(i,1)$ to insure that no other terms will interfere. C is some operator that will not be detailed here.

The factor $(1 + \delta)$ is used here as a safety factor to assure that the counterfactual relationships must hold, because otherwise a slight degradation in the amplitude advantage of the designated state would ruin the mapping; perhaps the same final brain states could receive amplitude from more than one initial state. All quantities are to be the appropriate ones for the given time. The projectors like $P_{bi}$ could cover enough phase space territory that they include the appropriate states for both times of interest, or they could themselves be time dependent in order to do so.

With the given example wavefunction, at time = 0, suppose that $|a| > (1 + \delta) |b|$, so the formal state would be $B_1$. Note that $|c|$ and $|d|$ don't come into play at this point, because of the projection onto the $|s_1\rangle$ subspace; they are in a different "world". One could also say that the first two terms of the example wavefunction represent different "worlds", but with this mapping they can interfere with each other.

The division between "near environment" $e_1$ and "distant environment" $s_1$ is somewhat arbitrary, but a particular mapping must allow counterfactuals, so there must be terms in the wavefunction such that, according to the mapping, the formal state would be different if those terms were different, as the second term in the example. There will also be many terms in the wavefunction (like the 3$^{rd}$ and 4$^{th}$ terms in the example) which the mapping pays no attention to. This is how it is possible simultaneously to take counterfactuals into account *and* for many other worlds to exist which may have a different brain state. The distant environment $s_1$ in this mapping must not be affected by which of the counterfactual brain states is the dominant one. On the other hand, slicing up the configuration space finer by putting $s_1$ closer at hand will help create more mappings, and the number of mappings must be maximized. Decoherence creates the suitable environments.

Why use expectation values or any kind of brackets in mappings? *Any* function of the physical states is a potential mapping, but using expectation values is a way to ensure that the transition rules for formal states will be satisfied. Other functions that obey the transition rules seem likely to lead to implementations that are of the same FCC and are not independent of those that do use brackets. The mapping need not use the largest amplitude to select the state, but most mappings with MCII#1 probably would since smaller terms would be more vulnerable to 'noise' and to interference from other mappings' variable counterfactual states within the branch.

If the dynamics are such that the counterfactuals will be satisfied, and the sub-states of the CSSA properly inherit independence from the mapping, the computation will be implemented. The inheritance of independence must come from the projection onto the $|b_i\rangle$ subspace for a formal state $B_i$. $B_i$ is shorthand for some formal structure, for example a set of N bits having particular values that together determine i; $i = i(c_1,\ldots,c_N)$. For each of these bits, the mapping must satisfy sub-state independence, established by the bit value assigned by the mapping not being affected by permutations of physical state labels that the other bits are affected by, as in the SONG example. For example, this would be satisfied if each bit depends on the positions of a distinct set of particles, the configuration of which could be interpreted as forming a switch (mechanical, electrical, hydrodynamic, a roomful of people manipulating Chinese symbols, etc.)

This provides one implementation of the computation, but there are many other mappings that would implement computations, and there can be no arbitrary choice as to how many to use. For each computation, a set of mappings must be chosen that maximizes the number of independent implementations, using a criterion such as MCII#1 or MCII#2. For evaluating probabilities, ratios of the number of implementations of different computations are needed, and a regularization is needed to evaluate this if the numbers are infinite.

The mapping above can be modified to yield families of many mappings. The families of interest will depend on the criterion used for independence of implementations.

MCII#1:

In this case, independent implementations must inherit independence from physical labels just like sub-states do. By slicing up configuration space as finely as the dynamics and counterfactuals will allow with mappings that are each based on a different region, the number of implementations can be maximized.

Let

$P'_{beij} = |be'_{ij}\rangle\langle be'_{ij}| \otimes \mathbb{1}_s \otimes \mathbb{1}_g$
$P'_{Sij} = \mathbb{1}_b \otimes \mathbb{1}_e \otimes (|s'_{ij}\rangle\langle s'_{ij}|) \otimes \mathbb{1}_g$
$P'_{gj} = \mathbb{1}_b \otimes \mathbb{1}_e \otimes \mathbb{1}_s \otimes |g_j\rangle\langle g_j|$

$|be'_{ij}\rangle$ is a smaller piece of the phase space region covered by $|b_i\rangle|e_i\rangle$, with the new index j to enumerate these pieces. Similarly, $|s'_{ij}\rangle$ is a smaller piece of $|s_i\rangle$. These finer-grained projectors replace the previous course ones: $F'(i,j,k,l,m) = \langle\Psi| P'_{beij} P'_{Skl} P'_{gm}|\Psi\rangle$ is then used to form mappings instead of $F(i,j)$.

The pieces involving the brain states must be chosen so that they still inherit independence within each mapping. There is more freedom with the $|s_{kl}\rangle$ and $|g_m\rangle$ regions. For example, if the index i represents a string of bits, then using states like $|B_1\rangle|e_1\rangle$ as the basis for the projector related to brain state 1, $|B_2\rangle|e_2\rangle$ for brain state 2, etc. would not generally allow inheritance because permuting the physical indices that generate inheritance of one bit would tend to result in states that the mapping does not cover, such as $|B_2\rangle|e_1\rangle$. It could however work if the proper inheritances are features of the $|e_i\rangle$ states.

If the regions used are too small, the implementations will fail because diffraction and dispersion would cause them to behave non-classically and also to interfere with each other. Each implementation must work given only the information in its own patch of phase space, regardless of what (within the limits of the mappings) is going on in the other patches. They can all share information about the parts of phase space that are outside of all of their sub-state differentiating regions.

If the wavefunction in one region has very high amplitude, then some effect is certain to spill over into neighboring regions. Each mapping must therefore include a limited range of amplitudes or limits on F'. This range, which can be different for different mappings, is chosen so as to maximize the actual number of runs of the computation given the actual wavefunction.

If the universe is infinite, the number of implementations can be infinite by slicing up the $|g\rangle$ region. $|be_{ij}\rangle|S_{kl}\rangle|g_m\rangle$ terms implement the computation, where the m-index has infinitely many possible values. A regularization is needed to compare probabilities, but the finite box regularization can suffice for that.

These implementations have been based on the brain states with the highest amplitude. Other brain states may implement computations with different mappings, but these are unlikely to be as robust, and thus these actual implementations will be less numerous.

The overall amplitude of the branch has not come into play in this analysis. However, near the edges of $|S_k\rangle$, fewer implementations may be possible due to spillover effects from other branches of the wavefunction (e.g. some $|S_{k'}\rangle$'s) or other sources of "noise". This could provide a way in which branch amplitudes do matter. The possibility of a "mangled worlds"-style derivation of the Born Rule from this effect will be investigated below.

MCII#2:

With MCII#2, like with MCII#1, more implementation mappings can be found based on the example mapping by using more narrow projections for each of them. In this case, however, the mappings are allowed to overlap in terms of the region of configuration space that they depend on. Instead, they must be independent in terms of setting up formal states by possible initial conditions, and the causal relationships must still hold with each mapping without interference from the others. The finer mappings must therefore be based on orthogonal projections.

Let $|s_i\rangle|g\rangle = \Sigma_j c_{ij} |sg'_{ij}\rangle$, where this is a sum of orthogonal terms. New mappings can be generated by using P'$_{sgij}$ = $\mathbb{1}_b \otimes \mathbb{1}_e \otimes (|sg'_{ij}\rangle\langle sg'_{ij}|)$ and F''$_{ijk}$ = $\langle\Psi| P_{bi} P'_{sgjk} |\Psi\rangle$ in place of F(i,j).

Cross-projections: Projection onto a term that includes multiple environment configurations (for example, $[c_i |s_i> + c_j |s_j>]|g>$) can also be usable for implementations because the proper causal relationships between formal states can be retained. Higher amplitudes could permit more implementations because even if the amplitude in the $|b_i>|s_i>$ projection is insufficient to dominate terms that favor other formal brain states, mixing in some of $|b_i>|s_i>$ in the projection could swing the balance. This possibility will be discussed later because it is related to overcoming 'noise' and Mangled Worlds.

Each orthogonal term would evolve in time independently, but orthogonal terms will always evolve into new terms that are orthogonal to each other at all times, since $<a(t)|b(t)>$ is conserved for any pair of time-dependent state kets $|a>$ and $|b>$. Therefore it will be possible to "read out" the final states by making appropriate projections.

There are infinitely many orthogonal terms that a function of a continuous variable can be divided into, so a regularization is needed to evaluate probabilities. Here, the Equal Norm Regularization Conjecture can be used, for example, to compare the probabilities of the runs of (for example) "$B_1$ then $B_3$" versus "$B_4$ then $B_6$" for the example wavefunction.

High energy states could be a problem. In dividing up a function on a finite region into a sum of infinitely many orthogonal functions, most of these will have to have high second derivatives, and therefore high kinetic energy (approaching infinity). Quantum gravity is likely to limit the number of such states possible with the finite region.

However, in an infinite universe, $|g>$ is not confined to a finite region. There could be an infinite number of $|sg'_{ij}>$ components of $|s_i>|g>$ that will serve without requiring very high energy states within a confined region.

If the ENRC is not used, and some finite box regularization is used instead, then once again amplitude seems to have no effect on probabilities, except for the effects of "noise" which could act as a low amplitude cutoff, giving higher amplitude branches more implementations.

**Derivation of Born's Rule from noise?**

Assume that at some *very* small amplitude level, phase space must be crowded with waves. The amplitude scale at which this happens will be called the State Crowding Amplitude Level (SCALe). Under these crowded conditions, computations become hard to implement for a wave at or below the SCALe. Counterfactuals get ruined because states can be affected by incoming waves; classical behavior does not emerge because decoherence can not set in without entanglement, and entanglement requires phase space to establish records in, while crowding phase space with amplitudes destroys records. The SCALe may be due to "random" noise or to more organized sources and may vary in different regions of phase space.

MCII#1:

Once again, it is important to recognize that due to the high dimensionality of phase space, most of the phase space volume in which a successful implementation mapping could be made is at the outer edges of the relevant regions – which is fortunate for attempts to derive the Born Rule because it is only near these borderline areas where the effect of the "noise" could possibly be large enough to matter. Such borders *must* exist, because otherwise amplitudes in any area in configuration space would be able to implement the computation, which is clearly false as the action of computers depends on their configuration. The effect of "noise" would be to shrink the border a little and to require more robust, and thus fewer, mappings just within the border.

It is crucial that brain size or structure not have much effect on the phase space volume in order for the scheme to work, since upon learning the outcome of an experiment, a person's brain would undergo structural changes such as synapse formation. One could argue as follows:

Suppose computer #1 relies on N bits to implement some FCC, while computer #2 relies on N+1 bits to implement some other FCC. Both computers are of identical construction, so that computer #1 has an extra bit of memory that it doesn't use. The measure of computer #1 is presumably larger that that of #2, because it is more robust – it can function in a larger volume of phase space, seemingly twice the volume, since it doesn't care if the extra bit is flipped. However, this will *not* be true of the phase volume that it will function in for an actual run, because if the amplitude of flipping is negligible, then so is the effective volume that is actually gained by not caring about the bit, because the borders of the low-measure flipped state would be relatively noisier compared to its low amplitude. If the situation gives substantial amplitude for flipping of the extra bit, that would decrease the amplitude of the unflipped state. So <u>if</u> the Born Rule would hold for computers if a constant number of bits of memory is always used, it would still hold if the number of bits used varies depending on the outcome seen.

There are three possibilities for the required "noise" within known physics:

1) The aggregate effects of the "tails" from *all* other branches of the wavefunction. This has some similarity to the Mangled Worlds Interpretation [Hanson] (see below).

Would this effect would be large enough to matter even near the edges of an implementation-supporting phase space region? There are many branches, but the tail effect from a branch in a distant part of configuration space would be very small, because it would mean that the distant branch has some amplitude for 'tunneling' into our branch. As pointed out by [Hanson], there would be a log-normal distribution of sub-branches for our branch, and many of those would have low amplitude. However, the noise may be smaller in those same sub-branches for the same reason; they are in parts of configuration space that are hard to reach with high amplitude. However, that may not be so; consider a sub-branch in which an unusually large number of nuclei have decayed, giving it low amplitude. This branch may be easier for noise to reach if the noise comes from a branch that started off with less of those nuclei.

In any case, there are limits to the region of configuration space that would support implementations of a particular FCC. The brain becomes entangled with objects in the environment, such as a ball. Consider a region of phase space in which the ball is displaced compared to where the brain thinks it is, perhaps due to sudden macroscopic tunneling of the ball. In such a region, the brain may still support the same FCC, because most of the amplitude in that region comes from the original branch, but the implementation will have to use very low wavefunction amplitudes. But as more and more changes to the environment are considered, making the environment resemble that of some other branch, at some point there will be more amplitude coming from the other branch to an alternative brain state than there is amplitude from the original branch to the original-FCC-granting brain state with that environmental configuration. At the cross-over point is the edge of the region that supports that FCC, so it is certain that 'noise' from the other branch will be important at that edge – that is what creates the edge.

The location of such an edge between branches i and j would be relatively insensitive to the amplitude ratio $A_i/A_j$ because amplitude falls exponentially in the number of degrees of freedom creating distance in configuration space, which is a much stronger factor.

At this point (pending more detailed study) it is reasonable to make some ansatz for the measure $M_i$ of the FCC supported by the brain in branch I, perhaps as follows:

$$M_i \approx \sum_j \frac{C_i A_i^2}{\varepsilon_{ij} A_j^2}$$

where
$A_i^2$ = the total squared amplitude of branch i
$C_i$ = a parameter based on the volume of the region and form of the tails
$\varepsilon_{ij}$ = a parameter based on the effect of noise from branch j on branch i along their edge, and the size of the edge, etc.
The case i=j accounts for any 'self-noise' and the diffraction limited volume
The sum is over all branches, not just whatever branches may have been created by the latest experiment.

This is not Born's Rule, which would be simply $M_i$ = constant · $A_i^2$.

There is at least a basic $A_i^2$ dependence here, so if for some reason the other factors are almost branch-independent, this might work. However the "noise" would be non-uniform (as the branches are in different parts of configuration space and affect each other) impacting on the probabilities. Regions of phase space that are unusual because few branches of any significant amplitude would be nearby would have enhanced probability compared to what the Born Rule predicts, because the number of implementations depends on the amplitude-to-noise ratio, and such regions would have low noise. This could be called the Rocket Problem because an example of it is that if a rocket is built and launched based on one spin-up measurement, and so such thing is done if the spin were down, the unusual action would seem to raise the probability for spin up.

On the other hand, most laboratory tests of Born's Rule do not involve such unusual macroscopic magnification of the effects of measurements, so the noise level for most of the outcomes might be nearly uniform, and this violation of the Born Rule *may* not have been experimentally falsified as of yet. The tentative conclusion is that this method fails.

2) Hawking radiation due to the cosmological constant populates the whole of configuration space with *very* small amplitude wavefunction "noise". This noise would be thermally uniform over phase space, so if it were the limiting factor it could eliminate the Rocket Problem. All states of a given energy are equally likely in a thermal distribution, and energy is conserved in quantum mechanical measurements [Mallah]. However, given how small this amplitude would be, ordinary tunneling from other branches of the wavefunction (as discussed above) seems more important.

3) The initial conditions of the cosmological wavefunction could imply the required low amplitude noise. It is well known that the initial conditions that would give rise to a universe like ours are of low entropy, as compared to a typical quantum mechanical state that a universe could be in, and that this gives rise to the thermodynamic arrow of time. If the initial conditions include the normal high amplitude/low entropy component, but also low amplitude components in all of phase space, the latter could supply the "noise". The advantage of this is that such "noise" could be distributed rather uniformly in configuration space (as in a thermal distribution), eliminating the Rocket Problem. This should give the Born Rule. The disadvantage is the appeal to unknown physics (the initial condition).

Next, this picture depends on the assumption that the "noise" will not itself give rise to too many conscious observers. (There could be some, as long as their measure does not overwhelm that of "normal" observers.) Despite the argument that a SCALe sets a lower limit for amplitudes to implement computations, the vast volume of phase space (though mostly in high entropy configurations) provides plenty of possibility for random noise to exhibit patches of organized behavior which could lead to "Boltzmann brains". It is difficult to compare these effects relative to "normal" observers with their aggregate amplitude tails and resulting many implementations, but this is definitely a problem that must be addressed for the "amplitude <u>only</u> matters to overcome noise" approach to be convincing.

The noise should not be correlated on scales appropriate for the aggregate mappings, if the derivation of the amplitude-squared measure is to be used (the $V^{1/2}$ dependence of total noise effect). This seems plausible if the aggregate mappings employ large aggregations.

Finally the proportion of amplitude-squared within the noise-limited tails should be branch-independent (at least to a good approximation). This is more plausible than the assumption that the noise itself is uniform, because very small, distant 'tails of the wavepacket' can result from events that are the same for all branches of interest.

The effects of entropy increase and passage of time must not be rapid exponential increase in the number of implementations if a noise-based derivation is to explain our observations. If the noise level remains constant, the number of branches may increase but there would have to be more aggregation of regions to overcome the noise, with the number of aggregates remaining proportional to amplitude-squared.

MCII#2:

"Cross-projection" could allow amplitude from the center of a wave-packet to assist the "edges" in overcoming interference. By placing the emphasis on available amplitude rather than physical regions, this may allow a better approximation to Born's Rule than MCII#1 would. More detailed study of the issues would be needed before drawing any firm conculsions.

**Comparison with Hanson's Mangled Worlds:**

[Hanson] created an interesting attempt to derive the Born Rule in the world-counting tradition. Due to the numerous microscopic scattering events that surely occur with unequal amplitudes for their various possible results, a distribution of fine-grained branches (leaving aside the question of precise definition of such) arises which has a log-normal dependence on amplitude, as it is a random walk in terms of multiplication of the original amplitude. Large amplitude branches are assumed to interfere with or 'mangle' smaller amplitude branches, imposing for all practical purposes a minimum amplitude cutoff. If the cutoff is in the right range of levels and is uniform for all branches, then due to the mathematical form of the log-normal function, the number of branches above the cutoff is proportional to the square of the original amplitude.

Unfortunately, this Mangled Worlds picture relies on many highly dubious assumptions, such as the uniformity of the 'mangling' cutoff. Nonetheless it brings to light some interesting facts, such as the tendency towards a log-normal distribution of "worlds".

Hanson used the density matrix formulation of QM in describing his approach. The density matrix formulation makes it easier to ignore the distant environment (which is assumed to be weakly coupled to the system of interest) except as a source of decoherence. Decoherence emerges by means of entanglement with the environment subspace. Such entanglement is not perfect, so decoherence is inexact.

In the the density matrix formulation, the environment subspace is traced over, forming a density matrix on the reduced phase space where the 'worlds' are defined. This provides a form of finite box regularization.

Hanson uses projections onto certain regions of phase space to represent 'worlds'. Consider a system of two 'worlds' L and S, so that world L could be represented by density matrix $\rho_{LL}$, and world S by $\rho_{SS}$; the system density matrix $\rho$ is

$$\rho = \begin{matrix} \rho_{LL} & \rho_{LS} \\ \rho_{SL} & \rho_{SS} \end{matrix}$$

with time evolution given by $i\hbar \, d/dt \, \rho = H\rho - \rho H + S$, where S describes interaction with the environment. In terms of the matrices $\rho_{LL}$ and $\rho_{SS}$:

$i\hbar \, d/dt \, \rho_{LL} = H_{LL} \rho_{LL} - \rho_{LL} H_{LL} + (H_{LS} \rho_{SL} - \rho_{LS} H_{SL}) + S_{LL}$

$i\hbar \, d/dt \, \rho_{SS} = H_{SS} \rho_{SS} - \rho_{SS} H_{SS} + (H_{LS} \rho_{SL} - \rho_{LS} H_{SL}) + S_{SS}$

$i\hbar \, d/dt \, \rho_{LS} = H_{LL} \rho_{LS} - \rho_{LL} H_{LS} + H_{LS} \rho_{SS} - \rho_{LS} H_{SS} + S_{LS}$

$i\hbar \, d/dt \, \rho_{SL} = H_{SL} \rho_{LL} - \rho_{SL} H_{LL} + H_{SS} \rho_{SL} - \rho_{SS} H_{SL} + S_{SL}$

The time evolution of 'worlds' L and S is described by the first two equations above. The terms in parentheses provide the effect of the other world on the given world.

Hanson assumes that the H operator terms are of similar magnitudes. However, if a term like $H_{LS}$ describes the coupling of two macroscopically distinguishable 'worlds', then such a term must be very small. It describes the transition of one macroscopically distinguishable state (such as a dead cat) into another (such as a living cat) by quantum 'tunneling'.

This is an important point because he assumes that if the amplitude of the 'small' world S is small enough compared to that of the large world, then even despite decoherence the time evolution of S will be dominated by the coupling term, causing S to be 'mangled'. If the coupling term is much smaller than he thought, then the ratio of these amplitudes would have to be correspondingly larger.

The ratio could be larger, but while the ratio can become thermodynamically large, it would be difficult to reach a large enough ratio without creating additional decoherence or further reducing the $H_{LS}$ terms.

There is another problem with the argument: Even if the ratio starts out large, if the $H_{LS}$ mediated transition is occuring, then after some time the ratio will no longer be so large.

Here computationalism may help some: "mangling" should be taken to mean that the small world will no longer implement the appropriate computation, or if it does, then it will have a much reduced number of implementations. If most of the amplitude in $\rho_{SS}$ is due to the $H_{LS}$ transition and similar transitions from other large 'worlds', then it is likely to be 'noisy' and not dominated by a single 'classical' brain state of greatest amplitude within the world S. Thus, S could be 'mangled' even if its amplitude is not small enough that the transition from larger worlds still dominates its current time evolution.

Hanson assumes that there is a uniform 'mangling' cutoff amplitude level, and if it falls in the right range, then the Born Rule follows (approximately) from world counting, given a log-normal distribution of world amplitudes. He admits that deviations from the Born Rule will appear to the extent that the cutoff level is not uniform. He suggests that a more 'global' source of mangling (a large world that affects many others) would produce a more uniform cutoff amplitude than if worlds that recently diverged affect each other more. Given the nature of the transition-driving terms like $H_{LS}$, however, more 'local' effects seem more likely, causing noticeable deviations from the Born Rule.

The 'mangled worlds' picture could be improved upon by taking a computationalist approach using MCII#2, treating the density matrix as a finite box regularization which is to be expanded out, and counting implementations instead of 'worlds'. If the number of implementations per 'world' is approximately constant due to the use of a finite-box regularization instead of the ENRC, the basic outlines of the argument could still follow. Instead of discarding 'mangled' worlds, however, they might be grouped together to overcome 'noise' which itself could give an amplitude-squared proportionality.

The basic problems would remain, however. $H_{LS}$ is too small to give mangling even for recently diverged worlds. As in the MCII#1 case, mappings near the 'borders' might overcome that issue, though the 'border' concept must be generalized. Even if that did work the 'Rocket Problem' of nonuniform cutoff would remain. With MCII#2, however, a more uniform 'cutoff' might for all practical purposes result from "cross-projection".

The number of implementations must not be a strong function of time. Even if noise is due to other branches, the noise level might not decrease with time, while the branches get more numerous but individually have less amplitude. Aggregation could work to overcome noise as described above. More detailed study would be needed to evaluate the feasibility of these suggestions.

**Hints of New Physics?**

The Born Rule may be compatible with computationalism within bare quantum mechanics, either by means of one of the above arguments or by some new approach. (As this paper is the first to introduce what is hopefully a robust enough criterion for sub-state independence, it seems reasonable to allow some time before all the implications and possibilities of computationalism can be considered as known.) But if in the end one concludes that the Born Rule is not compatible with computationalism within bare quantum mechanics, there are two choices: abandon computationalism, or modify the model of physics.

If computationalism would work fine for classical physics but is incompatible with bare quantum mechanics, the philosophical simplicity of computationalism is perhaps more important than the mathematical simplicity of the MWI. A need for modification of physics would degrade the simplicity and appeal of the MWI, but on the other hand, all other interpretations of QM already do modify physics and must still face the requirement to relate observers to the mathematical model.

1) The most obvious alternative model of physics to consider first is a hidden variables model such as the Pilot Wave Interpretation (PWI).  However, adding particle hidden variables to a Pilot Wave does nothing to remove the far more numerous wave implementations.  Since adding a single set of PWI-style hidden variables would have a completely negligable effect on observation probabilities with a computationalist approach, the PWI is ruled out as a meaningful option for the computationalist.

However, if an infinite number of sets of PWI-style hidden variables are added to the model it is a different story, since the implementations based on them would now outnumber those of the wavefunction alone.  This approach is taken in Continuum Bohmian Mechanics (CBM), in which instead of a particle guided by the Pilot Wave in configuration space, there is a continuous fluid or 'gas of dust'.  This is a Many Worlds Interpretation with hidden variables.  Since a distribution of PWI-style particles will evolve to approach that of the Born Rule, this model would indeed give the Born Rule within a computationalist approach.

The PWI requires a preferred reference frame, and CBM would too.  This is a disadvantage because it might end up incompatible with quantum gravity, but it is of course premature to declare it non-viable based on that speculation.  For the time being, CBM provides a possible fallback position for computationalism.  This should not, however, in any way discourage investigation into possible more elegant ways to reconcile computationalism with the Born Rule.

2)  Perhaps there is some actual small wavefunction amplitude cutoff (such as digital physics would bring). [Wolfram]  Or if wavefunctional values are digitized, perhaps ordinary noise is the limiting factor for normal observers, but the digital cutoff simply removes the ultra-tiny amplitudes that could otherwise infest the vast reaches of configuration space with Boltzmann Brains.  In either case such a cutoff would be so tiny that it could never be detected experimentally.

An actual 'resolution limit' for wavefunctions [Buniy] could have a similar effect to the Equal Norm Regularization Conjecture, except that it would not be just a mapping regularization but an actual physical limitation at some small but finite scale.

3) The All Universes Hypothesis (AUH) that all mathematically possible structures exist [Tegmark] has the problem of how it might select a unique and inevitable measure distribution.  If it can do so, however, it would be the simplest possible theory and thus strongly favored as the fundamental theory of physics.  It could also explain the existance of the universe in terms of the inevitable Platonic existance of mathematics.  In the digital sector, if a particular Turing machine is given all possible codes, the simpler ones (of shorter functional length) will more often be implemented as they are combined with more 'don't care' bits. (However, the choice of which Turing Machine to use appears arbitrary, giving even the digital sector a measure problem.)  It may be that a typical conscious observation would appear to be in a universe like ours, which may be implemented on some underlying simpler structure.  If that structure is digital, it could impose an effective small amplitude cutoff.

**Alternate philosophy of mind?**

If computationalism does not predict the Born Rule from the quantum wavefunction, one might abandon computationalism. Computationalism seems to be the best non-dualist approach (compared to the Activity Hypothesis for example), so a dualist approach might be tried.

Don Page's Sensible Quantum Mechanics [Page 2] provides an example of an alternative philosophy of mind that could be applied to QM. It is an explicitly dualist Many Worlds Interpretation, with psycho-physical laws in which projection operators applied to the wavefunction give the measure of various conscious observations.

The mathematical form of this may seem not so different from the use of mappings in computationalism, but the key difference is that no attempt is made to justify it as merely a precise characterization of the behavior of the underlying system, as computationalism does. Psycho-physical laws such as those suggested by Page would instead be new laws of nature that deal directly with consciousness. Because they are proposals for new laws of nature, the Born Rule can easily be inserted by hand, as Page does.

While there is thus no obstacle to deriving the Born Rule with such an approach, there is the drawback of philosophical implausibility, because it would imply that without these laws of nature, there would be no consciousness – people would be philosophical 'zombies'. Yet clearly, those of these 'zombies' that happen to experience the Born Rule would come to the exact same conclusions as people in our world do regardless of the law, since functioning (such as speech) is not at all affected by epiphenomenal consciousness, but only by what computations are implemented. There would also be no real explanation for why observations should closely match the physical world – it would be quite a 'lucky' break that the law of nature does that for us. (If it does; we couldn't write about what our own observations are since our brains don't take input from epiphenomena.) There would be no guarantee that artificial intelligences would be so 'lucky'. In addition, even some dualists [Chalmers 1] favor a computationalist approach due to its greater simplicity.

**Quantum Gravity and the possible "Frozen Formalism":**

Some attempts to model quantum gravity introduced the Wheeler-DeWitt (WDW) equation for a closed universe which gives rise to a "frozen formalism" in which there is no time evolution. Attempts have been made to derive an effective time by using some configurations as approximate clocks, but that would not suit the implementation criteria for computations outlined in this paper.

However, the "frozen formalism" approach has other problems. By eliminating time, the justification for deriving the WDW equation itself from temporal diffeomorphism invariance disappears. In addition, it would seem that the initial state of the universe should not be as *computationally deep* as the WDW equation would require. If Occam's

Razor can be applied to initial states, it is simpler to generate deep states by means of performing the computations instead of just having the state exist to begin with.

The problem can be eliminated either by postulating an open (infinite) universe, or by modifying the WDW equation to make it resemble a time-dependent Schrödinger equation, with non-zero energies allowed (and suitable initial conditions to resemble general relativity). [Weinstein] Another possibility to preserve computationalism even in the "frozen formalism" is to use some definition of counterfactuals suitable for degenerate frozen eigenstates.

Finally, it must be noted that the "frozen formalism" is probably the only way to save the single-world Pilot Wave Interpretation (PWI) within a computationalist framework. The hidden variables in that interpretation would not be frozen, so they could implement computations, while the wave function would be frozen and could not. (Without the frozen formalism, wave implementations would overwhelm those of a single-world set of hidden variables, turning the PWI for all practical purposes into the MWI.) However, the PWI's need of a preferred reference frame is violently at odds with the motivation for diffeomorphism invariance that inspires the frozen formalism.

**Conclusion:**

Computationalism is the least philosophically extravagant way that consciousness could be related to physics. Applying it to QM produces the MCI, a model which makes the MWI precise. While it is not clear that the Born Rule must follow from the MCI, it has been argued that under certain assumptions it could, such as with the use of MCII#2 + ENRC, MCII#3, or an approximate derivation based on 'noise' setting a for-all-practical-purposes minimum amplitude.

Three reasons were given why derivation of the Born Rule is important for the MWI, which can be summarized as 1) to show that the MWI need not imply some other wrong rule; 2) to avoid invoking dualism about consciousness; and 3) to retain the simplicity of the wavefunction-only dynamics.

If the MCI is compatible with the Born Rule, even if not truly providing a proof of the Born Rule, it would make these reasons irrelevant as objections to the MWI:

1) Compatibility with the Born Rule means that the MWI need not imply a wrong rule.
2) Computationalism is the best candidate for a theory of consciousness that avoids dualism. If even computationalism is taken to be dualistic, then dualism is unavoidable.
3) Adding the assumptions of computationalism does increase the complexity of the theory, but in a minimal way compared to other theories of consciousness. The complexity of the combined theory (the MWI + computationalism) is still small compared to the complexity of some other model of physics + some other model of conscious observers.

Even if computationalism is accepted, the assumptions about implementation counting needed to derive the Born Rule may not seem plausible. In that case, the most plausible models of physics are then either many-worlds hidden variables models such as Continuum Bohmian Mechanics, or digital physics imposing a small amplitude cutoff leading to a many worlds model of its own. More study of computationalism (especially regarding implementation independence) is certainly needed before any firm conclusions can be drawn about alternative models of physics.

**Glossary:**

Activity Hypothesis: As opposed to computationalism, this is the hypothesis that consciousness depends only on actual physical activity, not also on what the physical laws (transition rules) would imply about counterfactual states.

Boltzman Brain (BB): A brain which spontaneously arises due to random fluctuations. BBs present a problem to any model which relies on noise to obtain Born's Rule, because all of the vast configuration space must have some amplitude in it, and the BBs could arise in the appropriate locations if not ruined by the noise. Most observers could be BBs, which would be inconsistent with our observations. This problem could be overcome by imposing a digital amplitude cutoff (which could be smaller than that needed for simply obtaining the Born Rule due to a digital cutoff) or by showing that functioning BBs would be rarer than normal implementations. BBs are also a problem for the existence of a true cosmological constant [Page 1].

Border: The outer edges of a phase space region in which valid implementation mappings of a computation can be made. Border areas are the only areas where small amounts of noise could matter to the density of implementations, but due to the high dimensionality of phase space they should constitute most of the volume where such mappings are made.

Born Rule: The experimentally observed probabilities for outcomes of quantum measurements are proportional to the integrated squared amplitudes of the orthogonal wavefunction terms associated with the respective outcomes. Derivation of the Born Rule, which applies to observers, from the mathematical description of a quantum system is the key obstacle for the MWI. Many attempts other than the MCI have been made to *derive* the Born Rule from the MWI; it is argued in this paper that they all fail.

Combinatorial Structured State Automaton (CSSA): A computation characterized by the labelling structure of its states, which can be digital or analog. A wide variety of such structures are possible.

Computer: A physical system that implements computations. While any system does so, the term is usually only applied to systems which can implement a wide variety of interesting or useful computations.

Computation: A formal system with characteristic states and transition rules. A computation is a way of characterizing structure and function, and systems with the appropriate characteristics to behave like the computation are said to implement it.

Computationalism: The philosophical belief that consciousness arises as a result of implementation of computations by physical systems.

Configuration Space: The set of arguments which the wavefunction is a function of, except time. In nonrelativistic QM this is the set of classical particle configurations. The term 'phase space' is used as a synonym for configuration space in this paper, although classically it would also include the particle momenta.

Conscious Computation: A computation the implementation of which by a physical system would give rise to consciousness.

Continuous Collapse Models: Models such as GRW in which the wavefunction randomly localizes based on the Born Rule. However, the wavefunction still retains some small amplitude in all of configuration space, leaving these models in as much need of derivation of the Born Rule as the MWI.

Continuum Bohmian Mechanics (CBM): A Pilot Wave Interpretation in which instead of just one particle in configuration space there is a continuous fluid or gas. This is a realist MWI model which together with computationalism would give the Born Rule, but at the cost of being a speculative, more complex, and explicitly non-Lorentz-invariant model.

Counterfactuals: Computations have transition rules that specify what *would* happen given that the system were in any possible formal state, even though it is actually in just one state. These rules are an important part of the implementation criterea.

Digital Physics: Physics in which quantities such as the wavefunction can only take on a discrete set of values, rather than being continuous functions. This would imply a minimum amplitude and could help establish the Born Rule for the MWI, but at the cost of being a speculative, more complex, and explicitly non-Lorentz-invariant model.

Direct Derivation: If the number of implementations is proportional to the amplitude squared of even an isolated wavefunction without the need for noise or minimum amplitude, the Born Rule would be established directly for the MCI. This would avoid the Boltzmann Brain Problem. MCII#3 or another criterion might allow it.

Equal Norm Regularization Conjecture (ENRC): For use with MCII#2, the right regularization is conjectured to be one in which the number of mappings is proportional to amplitude-squared due to correspondence to orthogonal components (of state vectors) with equal norms. This is at odds with the possibly finite number of quantum states that may be available at any stage of a Finite Box Regularization.

Finite Box Regularization (FBR): Mappings in an infinite universe would require some regularization to allow evaluation of probabilities. The finite box regularization restricts the spatial volume that the mappings depend on at any stage of the regularization. The maximum entropy limit due to Quantum Gravity may imply that at any stage of the FBR there are only a finite number of quantum states available.

Fundamental Conscious Computation (FCC): A conscious computation that does not give rise to its consciousness by means of simulation of another conscious computation.

Implementation: The performance of a computation, which is done by any system with appropriate characteristics. In general, a system implements many computations, and a computation may be implemented multiple times over by the same system. Criterea are given in this paper for defining implementation and for counting multiple implementations.

Implementation Problem: If a traditional (naive) definition of implementation is used, most simple systems would implement every computation, which is not compatible with computationalism. A solution to the problem is proposed in this paper.

Inheritance: A computation can inherit part of its substate label structure from aspects of the system that implements it to help meet criterea for substate independence. This allows a function on several variables to implement a computation of several independent substates that depend on all parts of the function, but with restrictions.

Initial Conditions: The initial conditions of the universal wavefunctional could supply low amplitude noise that is uniform in phase space. This could help establish the Born Rule for the MCI without the Rocket Problem but at the cost of being a speculative model that appeals to as yet unknown physics.

Input: The transition rule for a computation need not specify the values of all substates for the next time step. The substates that are not specified take input because they can be changed by external factors, and will influence the subsequent course of the computation.

Locality: A desirable property of an interpretation of QM, in which activity in spacelike seperated regions has no effect on each other's measure or other properties. The Shrodinger picture MCI may be local or nonlocal depending on the MCII critereon chosen, but it is argued that nonlocal effects would be hard or impossible to observe. It is suggested that the Heisenberg picture [Rubin] should be used to produce a local MCI.

Log-normal distribution: The logarithm of a multiplicative random walk approaches a Gaussian distribution, and the amplitudes of branches of a wavefunction would be distributed this way due to splitting events of unequal amplitudes, as pointed out by [Hanson]. This may help in the derivation of the Born Rule for the MCI, if there is uniform minimum amplitude (an actual digital cutoff or noise-based effective cutoff) of the right magnitude.

Many Computations Interpretation (MCI): A version of the MWI in which observers are modelled as implementations of computations.

Many Worlds Interpretation (MWI): Any interpretation of quantum mechanics in which there are many actual observers that see different results of the same measurement. Some hidden variables interpretations are MWIs. When "the MWI" is referred to, it always means an interpretation with no hidden variables and no collapse of the wavefunction, in the tradition pioneered by Everett.

Mapping: A mapping labels some states of an underlying system as corresponding to formal states of a computation to be implemented. A mapping is *valid* if it satisfies all the needed criterea such as substate independence.

Measure: The quantity of consciousness or (in the discrete case) the number of observations, which observation 'probabilities' are proportional to. Total measure is not always conserved as a function of time.

Measure Counting Implementation Independence (MCII): Implementations that are not independent should not be counted seperately towards measure. Three alternatives are proposed for an independence critereon.

Measure Problem: In the MCI, the measure of a conscious observation is assumed to be proportional to the number of implementations that would give rise to that observation and are performed by a physical system. The philosophical problem of how to correctly specify how to count implementations arises.

Noise: Low amplitude, poorly correlated waves that could establish a SCALe. It is argued that relatively uncorrelated noise could help establish the Born Rule via a different mechanism than the usual for a small amplitude cutoff: the aggregation of phase space regions to overcome random noise for implementations at the borders of the regions that give rise to normal observers.

Pilot Wave Interpretation: A hidden variables model which adds a particle in configuration space to the wavefunction. The wavefunction would still implement its normal computations which would be far more numerous than the particle's, ruining the distinction between this model and the MWI for a computationalist approach.

Position Basis: It is assumed in this paper that the correct quantum mechanical basis in which to apply the criteria for implementation validity is the position basis. A fundamental basis is needed because otherwise the Implementation Problem could not be solved for a quantum system. For QFT, the basis used is the set of configurations of the particle fields with the fields considered as functions of position.

Probability: For a deterministic model like the MWI in which the states are assumed to be known, the only source of uncertainty is indexical; in other words, an observer does not know a priori which observer he is. Therefore the best rule to use to make predictions (or decisions for that matter) is the one that works best for the greatest number or quantity of observers. (Quantity of observers is called measure.) The role of the "probability of observing X" is therefore filled by the measure ratio (quantity of observations of "X") / (total quantity of observations) which is a deterministic replacement for the frequency of observing "X" which would classically be the long-run expected value of (number of cases that "X" is observed) / (total number of trials).

Regularization: If there are infinitely many implementations, a regularization of the mappings is needed to evaluate ratios, in which at any stage there are a finite number. The infinite case is approached as a limit. The choice of regularization can affect the probabilities (ratios) found, so a natural or philosophically justified choice is needed.

Rocket Problem: The noise due to all branches of the wavefunction would be non-uniform. If it is the limiting factor for the SCALe and is used as an effective small amplitude cutoff to derive observation probabilities, it would result in higher measure for observers in unusual situations such as being on a rocket. This violation of the Born Rule *might* not be easily detected in typical lab experiments, but it is highly suspect.

State Crowding Amplitude Level (SCALe): A low amplitude level at which most configurations have relatively significant wavefunction amplitude, thereby making it difficult or impossible for waves with such low amplitudes to implement the computations that would be performed by macroscopic computers, unless correlated over large regions. The effect on quantum mechanical observers could resemble that of a minimum amplitude (and could help establish the Born Rule) but without the need for digital physics.

Simulation: A computation which is implemented by another computation. A system that implements the underlying computation therefore must implement the simulation. An underlying computation that is itself implemented by something else is called a *virtual machine*.

Singly Occupied N-dimensional Grid (SONG): A function on an N-dimensional grid in which the function has a nonzero value at only one node at a time. A SONG is a prime example of a system in which mappings to formal states can inherit labels for SSI.

Size, Problem of: A larger or more flexible system could support an exponentially larger number of implementation mappings than a smaller or less flexible system, which is not plausible and certainly undesirable for social equality. However, these implementations would not be independent, so MCII criteria are proposed that would avoid this problem.

Sub-State: A state of a CSSA is a structured combination of substate variables. A typical example is a bit string, in which each bit is a substate of the string.

Sub-State Independence (SSI): SSI is required for a mapping to be valid. A criterion for SSI is proposed as a solution to the Implementation Problem.

[Page 2] Page, D. Mindless Sensationalism: A Quantum Framework for Consciousness. 2002. In "Consciousness: New Philosophical Essays". arXiv:quant-ph/0108039v1

Rubin, M. Locality in the Everett Interpretation of Quantum Field Theory. Found.Phys. 32 (2002) 1495-1523. arXiv:quant-ph/0204024v2

Searle, J.R. 1980. Minds, brains and programs. Behavioral and Brain Sciences 3:417-57

Tegmark, M. 2003. Parallel Universes. In "Science and Ultimate Reality: From Quantum to Cosmos". arXiv:astro-ph/0302131v1

Van Esch, P. On the Everett programme and the Born rule. arXiv:quant-ph/0505059 v1 9 May 2005

Wallace, D. Quantum Probability from Subjective Likelihood: improving on Deutsch's proof of the probability rule. arXiv:quant-ph/0312157v2 May 2005

Weinstein, M., and Akhoury, R. 2004. Cosmology Quantized in Cosmic Time. arXiv:hep-th/0406123v2

Wolfram, S. A New Kind of Science. Wolfram Media, Inc., May 14, 2002